\documentclass[10pt]{iopart}
\pdfoutput=1
\usepackage{graphicx}% Include figure files
\usepackage{dcolumn}% Align table columns on decimal point
\usepackage{bm}% bold math
\usepackage{hyperref}% add hypertext capabilities
\usepackage{gensymb}
\usepackage{blindtext}
\usepackage{cite}
\usepackage[normalem]{ulem} % Enables striking a horizontal line through text (\sout)
\usepackage{subcaption}
\graphicspath{ {figures/}}

\begin{document}
	\title[Influence of nanotube length and density on the plasmonic terahertz response of CNTs]{Influence of nanotube length and density on the plasmonic terahertz response of single-walled carbon nanotubes}
	\author{P Karlsen$^1$, M V Shuba$^2$, C Beckerleg$^1$, D I Yuko$^2$, P P Kuzhir$^2$, S A Maksimenko$^2$, V Ksenevich$^3$, Ho Viet$^3$, A G Nasibulin$^{4,5}$, R Tenne$^6$, and E Hendry$^1$}
	\address{$^1$ School of Physics, University of Exeter, Stocker Road, EX4 4QL, United Kingdom}
	\address{$^2$ Institute for Nuclear Problems, Belarus State University, Bobruiskaya 11, 220050 Minsk, Belarus}
	\address{$^3$ Department of Physics, Belarus State University, Nezalezhnastsi Avenue 4, 220030 Minsk, Belarus}
	\address{$^4$ Skolkovo Institute of Science and Technology, Skolkovo Innovation Center, Building 3, Moscow 143026, Russia}
	\address{$^5$ Department of Applied Physics, Aalto University, School of Science, P.O. Box 15100, FI-00076 Espoo, Finland}
	\address{$^6$ Department of Materials and Interfaces, Weizmann Institute of Science, Rehovot 76100, Israel}
	\eads{\mailto{peterkarlsen88@gmail.com}, \mailto{mikhail.shuba@gmail.com}}
	
	\begin{abstract}
		We measure the conductivity spectra of thin films comprising bundled single-walled carbon nanotubes (CNTs) of different average lengths in the frequency range 0.3--1000 THz and temperature interval 10--530 K. The observed temperature-induced changes in the terahertz conductivity spectra are shown to depend strongly on the average CNT length, with a conductivity around 1 THz that increases/decreases as the temperature increases for short/long tubes. This behaviour originates from the temperature dependence of the electron scattering rate, which we obtain from Drude fits of the measured conductivity in the range 0.3--2 THz for 10 $\mu$m length CNTs. This increasing scattering rate with temperature results in a subsequent broadening of the observed THz conductivity peak at higher temperatures and a shift to lower frequencies for increasing CNT length. Finally, we show that the change in conductivity with temperature depends not only on tube length, but also varies with tube density. We record the effective conductivities of composite films comprising mixtures of WS$_2$ nanotubes and CNTs vs CNT density for frequencies in the range 0.3--1 THz, finding that the conductivity increases/decreases for low/high density films as the temperature increases. This effect arises due to the density dependence of the effective length of conducting pathways in the composite films, which again leads to a shift and temperature dependent broadening of the THz conductivity peak. 		
	\end{abstract}
	
	\vspace{2pc}
	\noindent{\it Keywords}: Carbon Nanotubes, Terahertz, Dielectric Properties, Percolation, Temperature, Plasmon, Tungsten Disulfide 
	%\maketitle
	\ioptwocol

	\section{Introduction}
	Optical and electronic properties of single-walled carbon nanotubes (CNTs) have been under investigation for over two decades due to their fascinating physical properties and exciting potential for advanced applications \cite{Reich2008,Portnoi2008,He2014,Titova2015,Zubair2016}. Subsequent studies in the low-frequency, far infrared, terahertz and gigahertz ranges, led to proposing CNT-based composites as effective functional materials with tailored electromagnetic properties for these ranges \cite{Brosseau2012,Hartmann2014,Shuba2017}.
	
	One significant observation that has attracted much attention is a peak in the conductivity of CNT-based composites in the THz range, observed for the first time, to our knowledge, by Bommeli et al. \cite{Bommeli1997}. Two different mechanisms of its origin have been debated in literature. An interband transition corresponding to a small THz-range gap inherent to metallic CNTs \cite{Kane1997} (\textit{finite radius effect}) has been proposed as a possible mechanism of a non-Drude-like behaviour of CNTs in THz range \cite{Kampfrath2008,Ugawa1999}. An alternative mechanism has been proposed by Akima and Slepyan et al. \cite{Akima2006,Slepyan2006} as manifestation of the localized plasmon resonance in finite-length CNTs, which we denote the \textit{finite-length effect}. Theoretical modelling \cite{Slepyan2010a} and experimental observations \cite{Shuba2012,Zhang2013} substantiates the dominant role of the finite-length effect in the THz peak origin, at least at room temperature. While the two explanations are not mutually exclusive, understanding the true origin of this THz peak is key to a number of potential CNT applications \cite{Hartmann2014}. However, due to the inherent difficulty in fabricating isolated CNT samples, most measurements have been carried out on mixtures of CNTs with various distributions in length, density, thickness, chirality and bundle-size, and fabricated using a variety of techniques \cite{Kim2003,Kampfrath2008,Xu2009,Bauhofer2009,Slepyan2010a,Shuba2012,Wen2013}. Moreover, different temperature dependencies of the THz conductivity for CNTs  have been reported \cite{Borondics2006,Thirunavukkuarasu2010,Zhang2013,Morimoto2016} demonstrating the temperature dependence to be frequency dependent. Zhang et al. \cite{Zhang2013} shows a weak temperature variation in the CNT films conductivity below 3 THz, while the temperature has been shown to have a much more significant effect on the electronic transport properties of (i) nanotube composites at GHz frequencies \cite{Xu2007,Wen2013}, and (ii) CNT-films in the mid-infrared range \cite{Shuba2016}.
	
	In this paper, using terahertz time-domain spectroscopy  (THz-TDS), we investigate the influence of tube length and density on the temperature dependence of the THz conductivity spectra for thin-films comprising single-walled CNTs. Using films with different tube concentration we observe a plasmonic THz peak which is related to the effective length of conducting pathways in the CNT network. Due to this plasmonic behaviour, the frequency dependent conductivity is influenced by both tube length and density. This can in turn determine the response as a function of temperature, giving rise to a THz conductivity that increases and decreases with increasing temperature for the cases of low and high tube density, respectively.
	
	\section{Theoretical background}\label{sec:theory}	
	Throughout this paper, we choose to describe the frequency dependent electromagnetic response of the CNT-films in terms of the real part of the complex effective conductivity, Re$(\sigma(\nu))$, and real part of the relative permittivity, Re$(\epsilon(\nu))$, as in references\cite{Shuba2012,Ugawa1999a,Kampfrath2007a,Kampfrath2008,Slepyan2010a,Nishimura2007}. These are related through
	\begin{equation}\label{eq:cond}
	\epsilon(\nu)=1 + \frac{i\sigma(\nu)}{2\pi\nu\epsilon_0},
	\end{equation}
	where $\epsilon_0=8.85\times10^{-12}Fm^{-1}$ is the vacuum permittivity and $\nu$ is the frequency. We note that equation (\ref{eq:cond}) assumes an isotropic conducting dielectric material. Since our films consist of unaligned CNTs oriented mainly in the plane of the film, their in-plane effective conductivity and permittivity can be considered to be isotropic, with their relation described by equation (\ref{eq:cond}).
	
	We begin our theoretical analysis by considering how tube length is expected to influence the THz conductivity resonance (referred to here as the \textit{finite-length effect}). It is straightforward to show that tube length is expected to determine the peak frequency, as well as the temperature dependence of the conductivity spectra. Our carbon nanotubes samples (see Sec. \ref{sec:sampleprep}) contain mostly bundled CNTs, which are assembled into a percolating network. For now, we consider only a diluted composite containing isolated, randomly oriented CNT bundles. We will show that for this composite the temperature dependence of the THz conductivity spectra is affected by the finite-length effect in the CNTs.
	We use the Waterman-Truell formula to estimate the effective relative permittivity of a composite material \cite{Slepyan2010a} through
	\begin{eqnarray}\label{eq:WTF}
	\epsilon_{eff}(\nu)=\epsilon_h(\nu)+\frac{1}{3\epsilon_0}\sum_{j}\int_{0}^{\infty}\alpha_j(\nu,L)N_j(L)dL,
	\end{eqnarray}
	where the function $N_j(L)$ describes the number density of the CNT bundles of type j with radius $R_j$ and length $L$, and $\epsilon_h$ is the relative permittivity of the host material. The factor 1/3 in equation (\ref{eq:WTF}) is due to the random orientations of the inclusions; and $\alpha_j$ is the axial polarizability of CNT bundle, calculated using the integral equation approach \cite{Nemilentsau2010}. We note that equation (\ref{eq:WTF}) ignores the electromagnetic interactions between inclusions in the composites. The real part of the effective permittivity can be found as ${\rm Re}(\sigma_{eff})=2\pi\nu\epsilon_0{\rm Im}(\epsilon_{eff})$.
	
	In our calculations we use $\epsilon_h=1$ and the volume fraction occupied by the inclusions is $\phi=20\%$.  For simplicity we calculate the effective conductivity of a composite comprising identical bundles of either 1 $\mu$m or 10 $\mu$m in length. Each bundle consists of 7 tubes: two metallic tubes (12,0) and five semiconducting tubes (13,0) with energy gap of 0.816 eV and chemical potential of 0 eV. Here $(m,n)$ denotes the chiral indices of the tubes. The axial conductivity of the metallic CNT is approximated by the Drude formula \cite{Slepyan1999a}, defined by an electron relaxation time $\tau=1/\gamma$, where $\gamma$ is the electron scattering rate \cite{Slepyan2010a}. The THz conductivity of the semiconducting CNTs is supposed to be negligibly small and does not contribute to the THz response of the bundle.
	
\begin{figure}[tb]
		\centering
		\includegraphics[width=\linewidth]{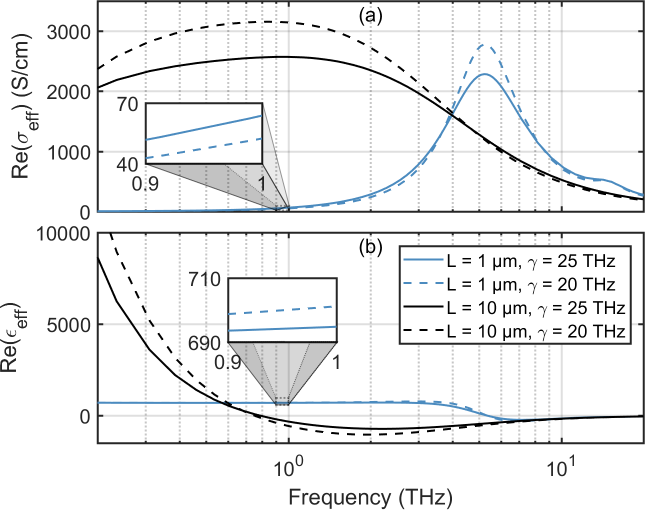}
		\caption{Calculated frequency dependence of (a) ${\rm Re}(\sigma_{eff})$ and (b) ${\rm Re}(\epsilon_{eff})$ at different carbon nanotube lengths $L$ and electron scattering rates $\gamma$. The insets show a zoomed area of their respective plots, highlighting the subtle differences for $L = 1$ $\mu$m when varying the scattering rate. }
		\label{fig:theoryeffeps}
	\end{figure}

	Figure \ref{fig:theoryeffeps} shows ${\rm Re}(\epsilon_{eff})$ and ${\rm Re}(\sigma_{eff})$ of the CNT composite for different lengths $L$ and electron scattering rates $\gamma$. The resonance at $5$ THz for $L = 1$ $\mu$m and the broad peak at $0.8$ THz for $L=10$ $\mu$m  are due to the well known finite-length effect in CNTs \cite{Slepyan2010a,Slepyan2006}.  Interestingly, the largest variation in both ${\rm Re}(\epsilon_{eff})$ and ${\rm Re}(\sigma_{eff})$ occurs in the locality of the THz resonance. In the range $\nu\in(0.1,1)$ THz, dependencies of the conductivity on the scattering rate for short and long tubes are different: both Re$(\epsilon_{eff})$ and ${\rm Re}(\sigma_{eff})$ decrease as $\gamma$ increases for long-length tubes ($L=10$ $\mu$m), while the opposite occurs for shorter tubes ($L=1$ $\mu$m) as highlighted by the insets. This behaviour arises due to the difference in response on and off resonance. 
	
	Since the dominant scattering mechanism for CNTs is expected to be acoustic phonon scattering\cite{Zhou2005a} (i.e. the scattering rate is expected to increase with increasing temperature, $T$) we can conclude that, if the effective conductivity is plasmonic in origin, its temperature dependence will be determined by CNT length. For long tubes (tens of microns) the finite-length effect is negligibly small in the THz range and the effective conductivity decreases with increasing temperature. This is expected for the intrinsic conductivity of an individual tube \cite{Zhou2005a}. For shorter tubes, the temperature dependence will be weaker and the conductivity may even increase with increasing temperature in the THz region. This behavior can be characterized by a crossover frequency, where the value of $\partial{\rm Re}( \sigma_{eff}) / \partial T$ changes sign. 
	
	As was recently shown \cite{Shuba2016}, the temperature induced modification of the infrared conductivity spectra of thin CNT films can occur due to the strong temperature dependence of unintentionally doped semiconducting tubes with Fermi energy located close to the top of the valence band. To demonstrate possible contribution of these tubes to the THz conductivity spectra, we calculated the same CNT-composite as in figure \ref{fig:theoryeffeps}, but assuming that (13,0) CNTs with energy gap of $0.816$ eV has a chemical potential of $0.4$ eV and electron scattering rate for all the tubes in the bundle is $\gamma=25$ THz.  For this case the static conductivity of the semiconducting CNTs, calculated using the model presented in \cite{Shuba2016}, is about 8 times smaller than that of metallic CNTs. 
	
\begin{figure}[tb]
		\centering
		\includegraphics[width=\linewidth]{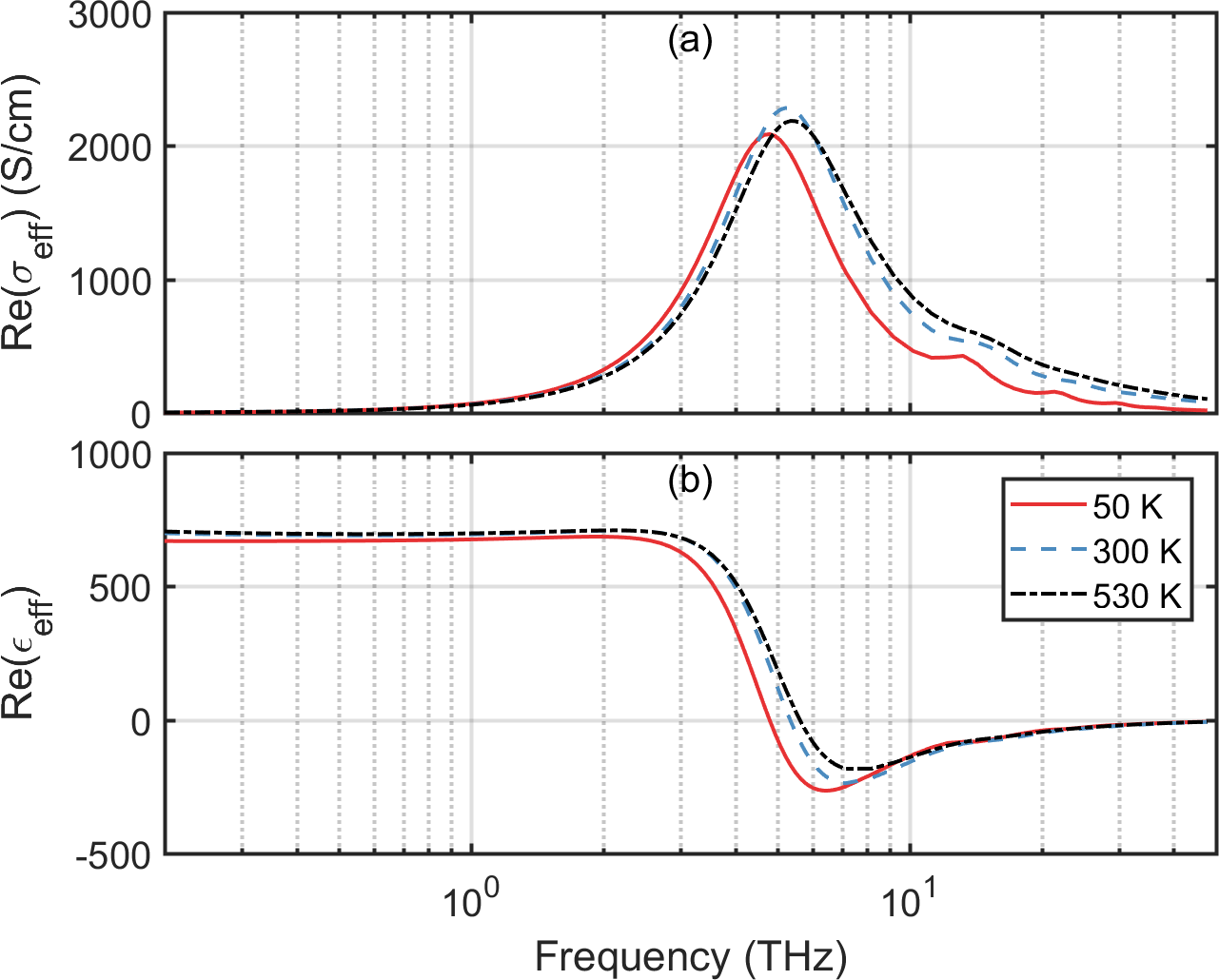}
		\caption{Calculated frequency dependence of (a) ${\rm Re}(\sigma_{eff})$, and (b) the ${\rm Re}(\epsilon_{eff})$ at 50, 300, and 530 K for the composite medium with bundled CNTs. Each bundle comprises doped semiconducting tubes.}
		\label{fig:theorybundle}
	\end{figure}

	In figure \ref{fig:theorybundle}, we present the values of ${\rm Re}(\epsilon_{eff})$ and ${\rm Re}(\sigma_{eff})$ for a composite containing identical bundles of $1$ $\mu$m in length at 50 K, 300 K, and 530 K, i.e. the temperature induced change in the conductivity is due to varying the electron Fermi-distribution function of the system. We note that in reality the scattering rate is expected to increase with temperature\cite{Kane1998,Zhou2005}, however in this case we choose a single scattering rate to show the effect of the Fermi distribution. One can see in figure \ref{fig:theorybundle} that the THz peak is expected to blue-shift slightly on increasing the temperature. In absence of a temperature dependent scattering rate, this effect arises due to an increase of the conductivity of the semiconducting tubes with temperature and, consequently, to the increase of the surface wave velocity in the CNT bundle\cite{Shuba2007}. The model predicts no temperature induced shift of the THz peak for composite comprising individual doped CNTs. This is discussed in section \ref{sec:exp} and confirmed in the supplementary material section S1, where we compare the data for bundled and unbundled CNT samples.
	Similar results were claimed in\cite{Nemilentsau2010} with respect to the doping effect: substitutional doping leads to the blue-shift of the THz conductivity peak for a composite comprising bundled tubes; there is no shift if tubes are individual.
	
	\section{Sample preparation}\label{sec:sampleprep}
	To study the influence of tube length on the temperature induced modification of THz conductivity spectra we prepared three types of films comprising CNTs in bundled form, where the average lengths of the CNT bundles in the films varies significantly: 
	
	(i) Free-standing films comprising long length single-walled CNTs ($l$-CNT, $L\approx 10$ $\mu$m, film thickness of 42 nm) with diameters of approximately 1.6 nm was prepared via the aerosol chemical vapour deposition method \cite{Nasibulin2011,Tian2011}. 
	
	(ii) Free standing films comprising medium length single-walled CNTs ($m$-CNT, $L\in(0.3,2)$ $\mu$m, average length of about 1  $\mu$m, film thickness of 500 nm) were prepared via the vacuum filtration technique \cite{Hennrich2002}. Briefly, a material of non-purified High Pressure Carbon Monoxide (HiPco) CNTs with diameters of 0.8--1.2 nm ({NanoIntegris} Inc.) were dispersed by ultrasonic treatment (Ultrasonic device UZDN-2T, 44 kHz, maximum power) for 1 hour in an aqueous suspension with 1\% Sodium-Dodecyl-Sulfate (SDS). Ultrasonic treatment cut the initially long tubes down to a length of up to 1--2 $\mu$m \cite{Hennrich2007}. Then, the suspension was centrifuged at 10000g for 15 minutes. Strong centrifugation leads to purification of the tubes and removes aggregated tubes. The suspension was then filtrated through a membrane, causing a film to collect on the filter. This film was then washed to remove all surfactant. Finally, the filter paper was dissolved by acetone and the film was transferred on to a  metallic frame with a hole of 8 mm in diameter.
	
	(iii) Thin films of short length single-walled CNTs ($s$-CNT, $L<300$nm, film thickness of 600 nm) were also prepared via the vacuum filtration technique. To obtain short tubes, highly purified (99\%) HiPco CNTs with diameters of 0.8--1.2nm ({NanoIntegris} Inc.), were cut by ultrasonic treatment of the material in a mixture of nitric and sulfuric acids \cite{Shuba2012N}.    
	
	Since the observation of a conductivity threshold in polymer/carbon nanotube composites \cite{Coleman1998} much research has been dedicated to fabricating CNT-based composite materials with percolated networks and understanding their optical and electrical responses  \cite{Bauhofer2009}. An ongoing issue with preparing CNT composite materials is aggregation, since tubes tend to be electrostatically attracted to each other instead of dispersing uniformly throughout the host material. To study the influence of the tube density on the temperature induced modification of the THz conductivity, we prepared a hybrid composite material consisting of well-dispersed and non-aggregated $m$-CNTs mixed with non-conductive inorganic WS$_{2}$ nanotubes (INT). INTs were synthesized in the large scale fluidized bed reactor (see details in \cite{Zak2009}). INTs have diameters 20--180 nm and lengths 1--10 $\mu$m; they are semiconducting with a bandgap of $2$ eV, and are transparent and non-conductive for THz frequencies. INT material was dispersed in 1\% SDS aqueous solution for $30$ min by ultrasonication and then the suspension was immediately centrifugated for 15 min at 300g to remove aggregated INTs. The suspensions of INTs and $m$-CNTs were mixed in different proportions and then filtrated to obtain thin films with thickness between $0.7$ and $12$ $\mu$m.  The quantity of both $m$-CNTs and INTs in suspensions before mixing was controlled via ultraviolet-visible spectroscopy (RV2201 spectrophotometer, ZAO SOLAR, Belarus). This provided us with films of different volume fractions of CNTs. The films were transferred to 10 $\mu$m thick PTFE and 1 mm thick quartz substrates for terahertz and microwave measurements. Finally the films on quartz substrates were annealed at 500\degree{}C for 30 minutes.  Film thickness was measured with the profilometer Veeco Dektak 6 M. 
	
\begin{figure}[tb]
		\centering
		\includegraphics[width=\linewidth]{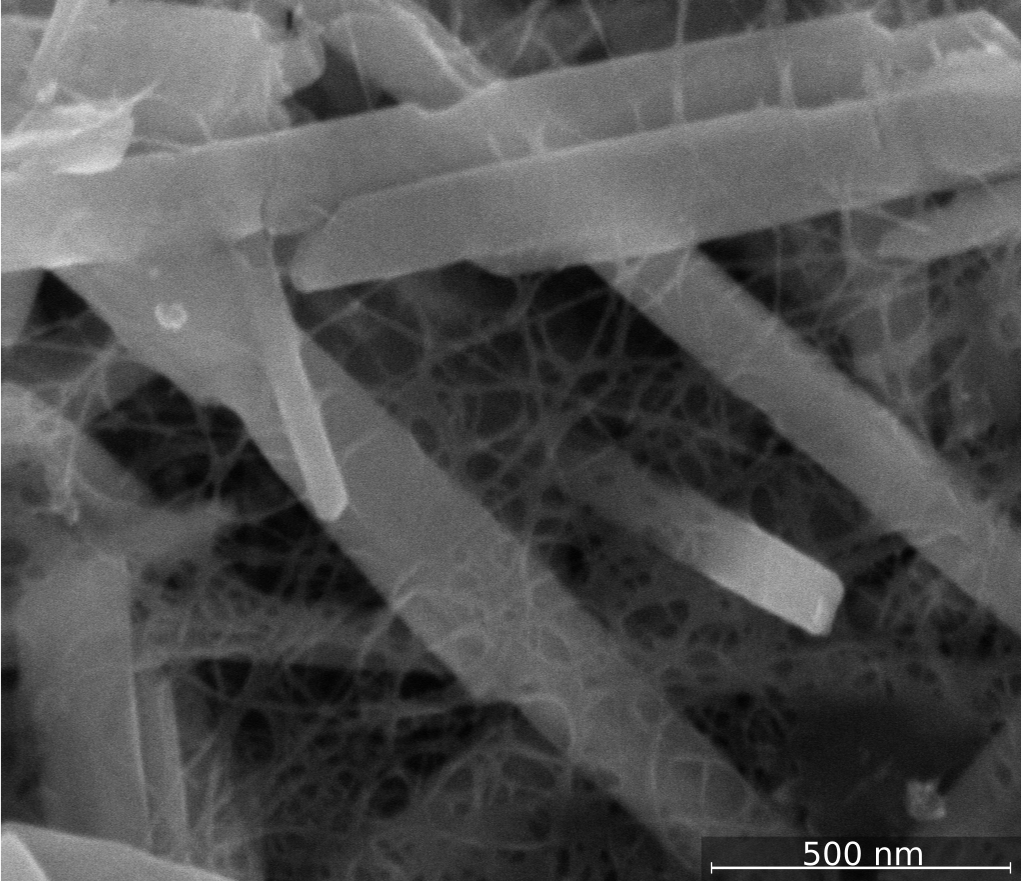}
		\caption{SEM image of samples comprising INTs and $m$-CNTs. The thick white tubes are INTs and the thin white curves are the CNT network. Note that the CNTs are well dispersed among the INTs.}
		\label{fig:SEM}
	\end{figure}

	The prepared CNT/INT composite samples were investigated using a Scanning Electron Microscope (SEM), see figure \ref{fig:SEM}: this clearly demonstrates the successful dispersion and non-aggregation of the CNTs, since the nanotubes are dispersed evenly in the sample and no dense particles comprising ten or more nanotubes can be observed. Finally, in supplementary material section S1 we compare the bundled $m$-CNT/INT samples with similar hybrid samples of unbundled Arch-Discharge single-walled CNTs and show that terahertz conductivity of the composite is higher for individual than bundled CNTs.
	To estimate the crystalline quality of the CNT material, Raman spectroscopy was applied (see supplementary material section S2). Raman spectra were obtained using a Raman spectrometer combined with a confocal microscope Nanofinder High End (Tokyo Instruments) at an excitation wavelength of 532nm.
	
	\section{Experimental measurement and analysis}
	With normally incident light, the optical density spectra were obtained in the ranges 0.1--2 THz, 2--270 THz, and 270--1000 THz  using THz-TDS (EKSPLA, Lithuania), a  Fourier-transform infrared spectrometer Vertex 70 (Bruker), and an RV2201 spectrophotometer (ZAO SOLAR, Belarus), respectively, see supplementary material section S3 for details. Data obtained in the range 1.5--3 THz has an error of about $20$\% because of high level of noises.
	
	To investigate the temperature dependence of the THz conductivity in the range 10--300 K, we employed a Terahertz Time-Domain Spectroscopy (THz-TDS) setup with access to a closed cycle helium cryostat (\href{http://www.arscryo.com/}{ARS})\cite{Ulbricht2011}.  THz pulses with a bandwidth of 0.3--2 THz were generated and detected by commercially available Photoconductive Antennas (PCAs) from \href{http://www.batop.com/index.html}{Batop} using a 40 MHz, 1064nm, femtosecond fibre laser from \href{http://ekspla.com/}{Ekspla}. From direct measurement of the complex transmission coefficient of our samples, the conductivity and dielectric function was obtained using (\ref{eq:fresnel}) (see below).
	
	The same THz measurements in the range 300--530 K were done in air using a home-made furnace-style heating sample-holder. In order to remove doping effects from air molecules such as $H_2O$ and $O_2$, which are non-covalently attached to the CNT walls, each sample was maintained at temperature of 530 K for 10 min resulting in a reduction of sample conductivity of approximately 20\%, before starting the measurements, and subsequently cooled from 530 K to 300 K while performing the THz measurements. We note that the sample conductivity was restored in one day after the heating due to the adsorption of air molecule on the CNTs.  
	The microwave measurements at 30 GHz were performed at room temperature by waveguide method\cite{GHz,Chung2007,Paddubskaya2016} using a scalar network analyzer R2-408R (ELMIKA, Vilnius, Lithuania). Static electrical conductivity of thin films was measured using a four-point linear probe technique. 
	
	One method for determining the permittivity experimentally is via the complex transmission function $t(\nu)$, which can be determined indirectly from transmittance spectra via the Kramer-Kronig relations \cite{Zhang2013}, or directly from time-resolved measurements such as THz-TDS. By measuring the electric field $E$ transmitted through the sample and a corresponding reference field $E_{ref}$ (in our case, either the quartz substrate alone, $\epsilon_{quartz}=1.96^2$, or air, $\epsilon_{air}=1$ in the case of free-standing films), the permittivity of the sample can be obtained from its transmission coefficient $t(\nu)=E/{E_{ref}}$ by solving the Fresnel equation of the system\cite{Ulbricht2011}:
	\begin{equation}\label{eq:fresnel}
	t=\frac{t_{12}t_{23}\exp(2i\pi\nu d_2(\sqrt{\epsilon_2}-\sqrt{\epsilon_1})/c)}{t_{13}(1-r_{21}r_{23}\exp(4i\pi\nu \sqrt{\epsilon_2}d_2/c))},
	\end{equation}
	where $t_{ij}=2\sqrt{\epsilon_i}/(\sqrt{\epsilon_i}+\sqrt{\epsilon_j})$ and $r_{ij}=(\sqrt{\epsilon_i}-\sqrt{\epsilon_j})/(\sqrt{\epsilon_i}+\sqrt{\epsilon_j})$ are the Fresnel transmission and reflection coefficients for normal incidence, $\epsilon_i$ is the permittivity and $d_i$ is thickness of region $i$. In our case, region 1 and 3 are the incident and transmitted regions, respectively, and region 2 is the CNT thin-film.
	
	\section{Experimental Results and Discussion}\label{sec:exp}
	\subsection{The finite-length effect}
	While many groups have observed a broad THz conductivity peak in CNTs, its origin has been debated for some time\cite{Kampfrath2008,PhysRevB.60.R11305,Zhang2013,Slepyan2010a,Shuba2012}. The peak was claimed to be associated either with a small-gap interband transition \cite{Kampfrath2008,PhysRevB.60.R11305} or with a localized plasmonic resonance\cite{Zhang2013,Slepyan2010a,Shuba2012,Slepyan2006}. However, both mechanisms contribute simultaneously and cause the appearance only one terahertz peak\cite{Slepyan2010a}. As was shown in\cite{Slepyan2010a}, the contribution of the finite-length effect is dominant at room temperature, when the gap energy is less than or comparable to the average thermal energy. The dependence of the THz peak frequency on CNT length has been experimentally shown in\cite{Shuba2012}, giving clear evidence of a plasmonic resonance in these materials. 	
	
\begin{figure}[tb]
		\centering
		\includegraphics[width=1\linewidth]{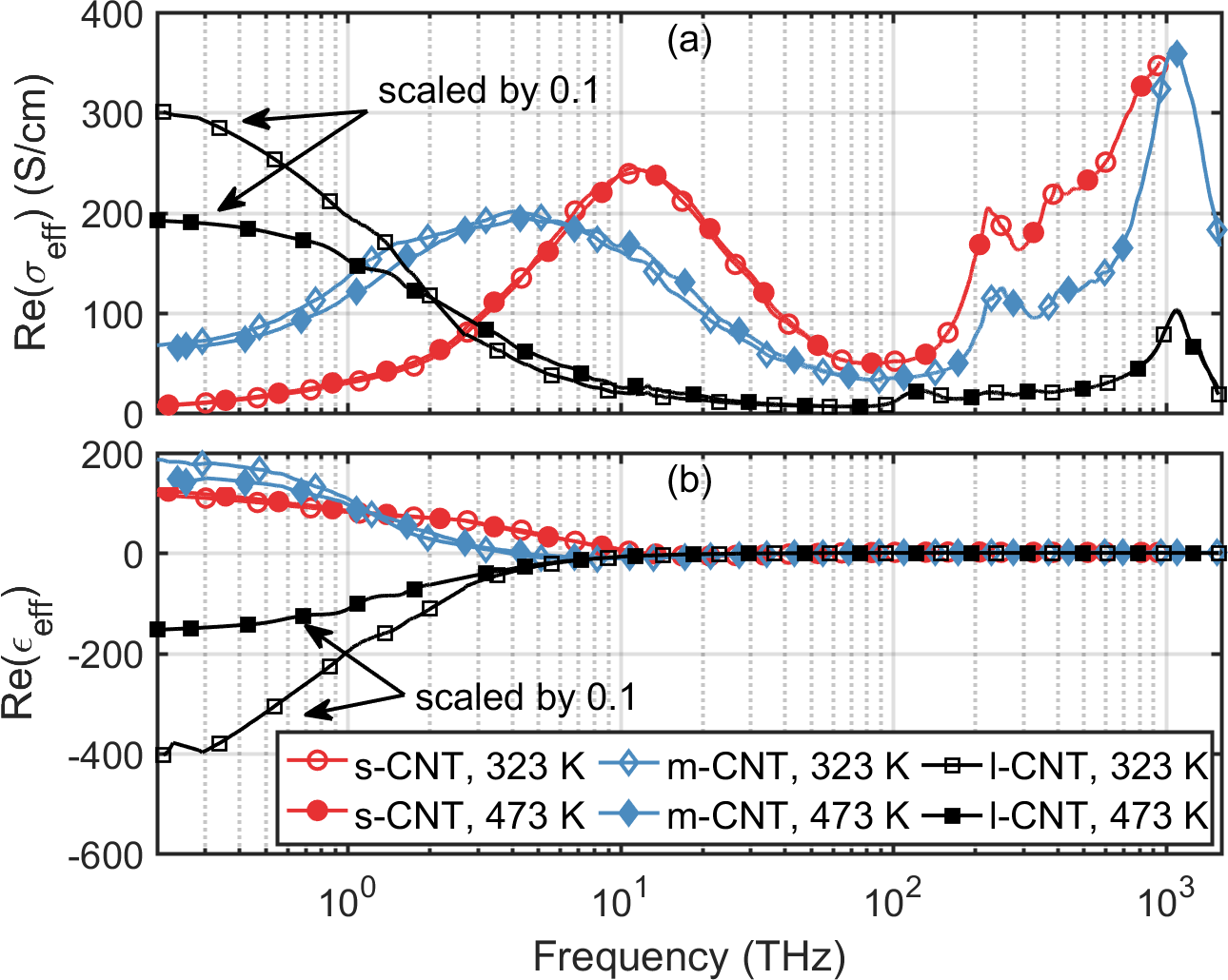}
		\caption{Frequency dependence of (a) the real part of the effective conductivity Re$(\sigma_{eff})$, and (b) the real part of the effective permittivity Re$(\epsilon_{eff})$, obtained from broadband optical density measurements at 323 K and 473 K for $s$-, $m$-, and $l$-CNT films. Note that the data for the $l$-CNTs have been scaled by 0.1.}
		\label{fig:od}
	\end{figure}

	In the present section we demonstrate experimentally the influence of the tube length on the temperature dependence of the THz conductivity spectra for CNT thin films. Good qualitative agreement of the experimental results with theoretical predictions made in section \ref{sec:theory} confirm the plasmonic nature of the CNT response in the terahertz range. Figure \ref{fig:od} shows conductivity spectra of the real part of the effective conductivity Re$(\sigma_{eff})$ and the real part of the effective permittivity Re$(\epsilon_{eff})$ measured at 323 K and 473 K for $s$-, $m$-, and $l$-CNT films. The optical density spectra for these films are shown in supplementary material section S1. As shown in figure \ref{fig:od}, the spectra for $m$- and $s$-CNT films have a broad peak at 4 THz and 10 THz, respectively. The spectra of $l$-CNT films demonstrates Drude-like behaviour.
	
	We note that the average diameter of $l$-CNTs differs from that of $m$-, and $s$-CNTs. It has been demonstrated previously that the polarizability and conductance of an individual single-walled CNT slightly depends on its diameter in the frequency range below interband transitions \cite{Slepyan2010a}. In terms of the effective conductivity of the CNT film, the tube diameter can influence this via intertube tunnelling or by varying the tube number density in the samples, however its contribution to the frequency of the terahertz peak is much smaller than from the finite-length effect. This was shown experimentally for samples with different average tube diameters of 0.8, 1, and 1.4 nm \cite{Shuba2012}.
	
	The temperature dependencies of the calculated spectra in figure \ref{fig:theoryeffeps} and measured spectra in figure \ref{fig:od} are very similar. Namely, in the range 0.3--1 THz, (i) long tubes $l$-CNTs ($L\approx 10 \mu$m) demonstrate stronger temperature dependence than shorter tubes $m$- and $s$-CNTs ($L< 2 \mu$m), (ii) The conductivity of $l$- and $m$-CNTs  decreases with increasing temperature, while the opposite is true  for $s$-CNTs. This can be explained by a decrease of the electron scattering rate with increasing temperature. As predicted from the model of localized plasmon resonance in figure \ref{fig:theoryeffeps}, the THz peak shifts to higher frequencies as the length decreases, resulting in a weakening of the temperature response.  
	
	As shown in figure \ref{fig:od} for $m$- and $s$-CNTs, the THz peak shifts to a slightly higher frequency with increasing temperature. We associate this small temperature-induced shift with the temperature dependence of the conductivity of some fraction of the doped semiconducting tubes. This effect was demonstrated in figure \ref{fig:theorybundle} of section \ref{sec:theory} for composite comprising bundled tubes. As follows from the model, this shift does not happen for individual tubes. This is supported by our measurements conducted for individual electronically separated tubes (see supplementary material section S1).  
	
\begin{figure}[tb]
		\centering
		\includegraphics[width=1\linewidth]{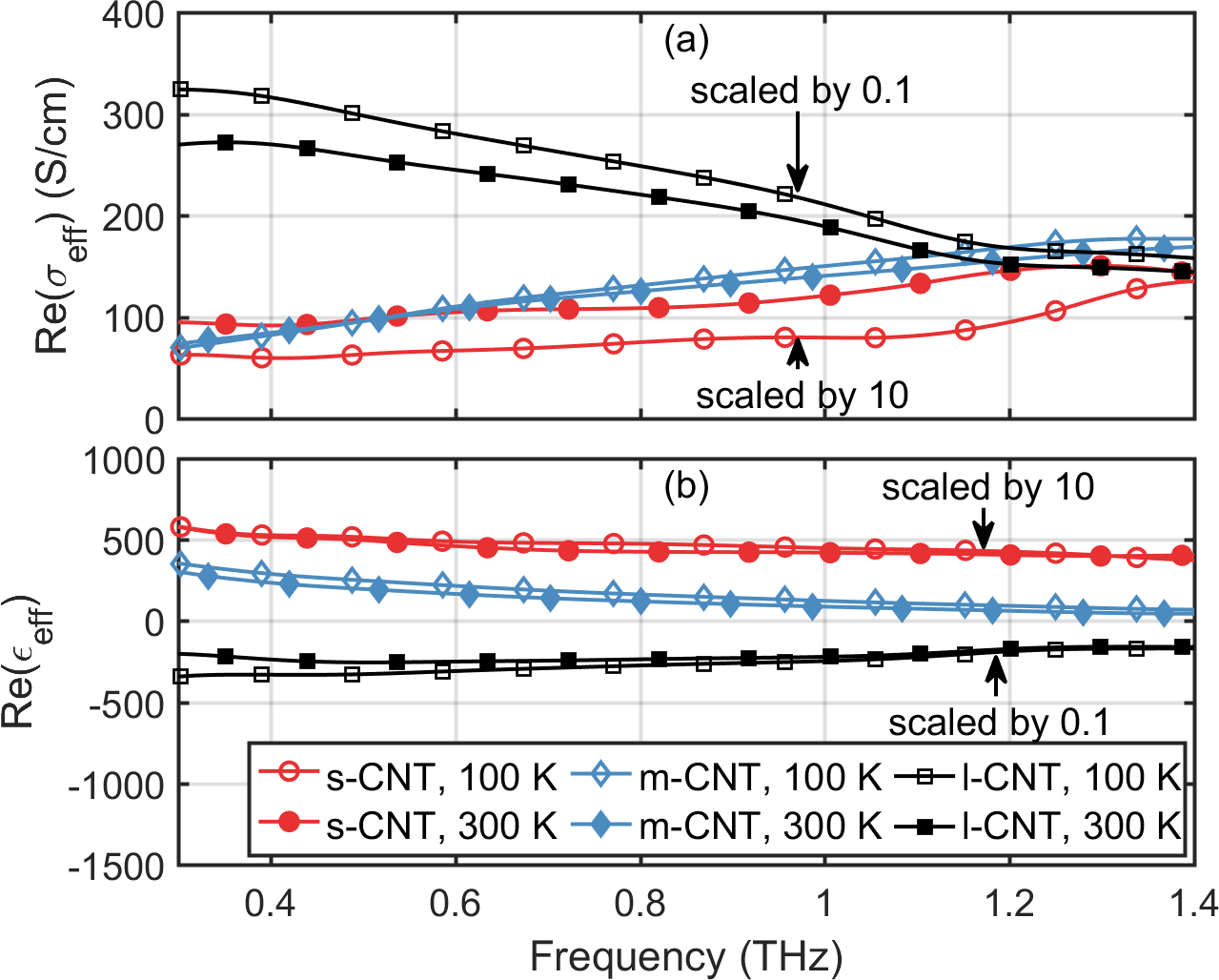}
		\caption{Frequency dependence of (a) Re$(\sigma_{eff})$, and (b) Re$(\epsilon_{eff})$, obtained from THz-TDS measurements of $s$-, $m$-, and $l$-CNT films at 100 K and 300 K. Note that the data for $l$- and $s$-CNT have been scaled by 0.1 and 10, respectively.} \label{fig:cond_low_temp}
	\end{figure}

	Figure \ref{fig:cond_low_temp} shows spectra of Re$(\sigma_{eff})$ and  Re$(\epsilon_{eff})$ measured for $s$-, $m$-, and $l$-CNT films at low temperature 100 K and 300 K. The behaviour of the conductivity and permittivity for these films are qualitatively the same as in high temperature range 300--530 K (figure \ref{fig:od}) and agrees well with the predicted finite-length response in figure \ref{fig:theoryeffeps} showing opposite temperature dependencies for short and long CNTs.
	
	\subsection{Electron scattering rate in long-length CNTs}
	In general, the THz conductivity spectra of CNT films are fitted with a Drude-Lorentz function containing the Drude and Lorentz terms with two and three unknown parameters, respectively\cite{Borondics2006,Zhang2013}. 
	The Drude term describes the contribution from ``infinite" length tubes while the Lorentz term is due to the localized plasmon resonance in finite-length CNTs. If the Drude- and Lorentz terms are overlapped with each other, some ambiguity in the determination of unknown parameters appears in the fitting procedure. If the tube length is large enough then the contribution from the Drude term dominates and we can fit the terahertz conductivity of the CNT film using just Drude term with two unknown parameters: the plasma frequency and electron scattering rate.  For this case, the effective conductivity of the film is a sum of the intrinsic tube conductance in the unit volume, with a factor responsible for orientation of CNTs in the film. In this approximation we can conclude that the electron scattering rate in the Drude term of the CNT film effective conductivity is equal to the electron scattering rate in the individual CNTs. Our $l$-CNT sample comprises mainly long tubes and, consequently, demonstrates Drude-like behaviour in the terahertz range.
	
	From our Drude fits of the effective conductivity for $l$-CNT samples in the range 0.3--2 THz at different temperatures, we found that the plasma frequency remains temperature independent, while the electron scattering rate $\gamma$ shows a clear temperature dependence. This result indicates that the temperature dependence of the intrinsic CNT conductivity is not related to the energy dependent density of states of the CNT, but can instead be attributed to temperature dependent electron scattering processes. Raman spectra of the $l$-CNTs show a very small D-mode (see supplementary material section S2), demonstrating a low defect density in the $l$-CNTs. Thus, we shall omit possible electron scattering by defects in our further consideration. 
	
\begin{figure}[tb]
		\centering
		\includegraphics[width=1\linewidth]{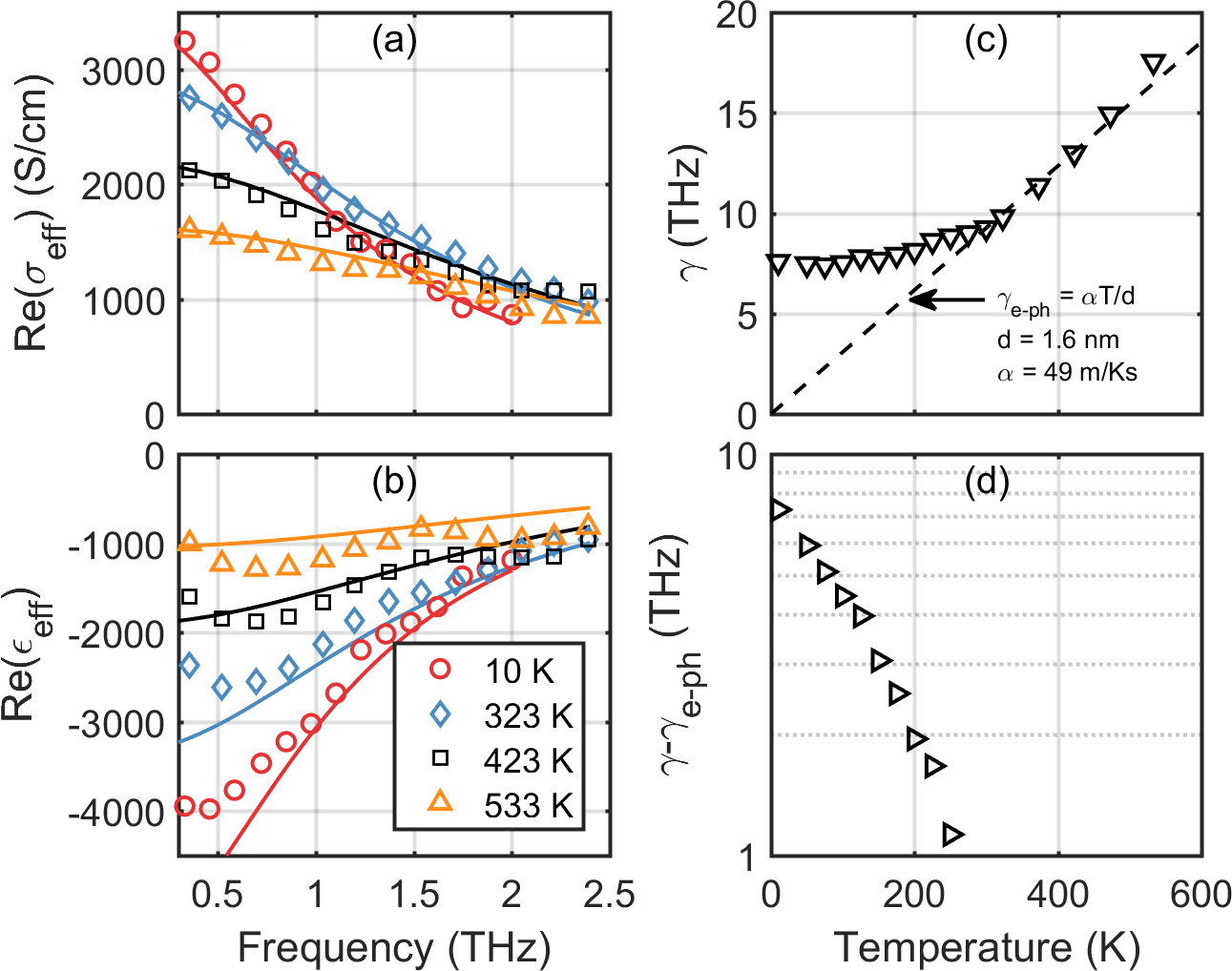}
		\caption{Frequency dependence of (a) Re$(\sigma_{eff})$, and (b) Re$(\epsilon_{eff})$, obtained from THz-TDS measurements of $l$-CNT film at 10 K, 323 K, 423 K and 533 K, along with their associated Drude fits (\full). (c) Temperature dependence of the scattering rate $\gamma$ (\opentriangledown) obtained from Drude fits of measured spectra Re$(\sigma_{eff}(\omega))$ and Re$(\epsilon_{eff}(\omega))$ in the range 0.3--2 THz for $l$-CNT film. The data has been fitted in the range 300--500 K with a linear function $\gamma_{e-ph}=\alpha T/d$ (\dashed), where d = 1.6 nm is the mean diameter of our CNTs, and $\alpha=49$ m/Ks is a fitting constant. (d) Temperature dependence of the value $\gamma-\gamma_{e-ph}$ calculated from the data in (c).} \label{fig:scat_rate}
	\end{figure}

	Figure \ref{fig:scat_rate}a and \ref{fig:scat_rate}b shows experimental and associated Drude-fits of the temperature dependence of Re$(\sigma_{eff})$ and Re$(\epsilon_{eff})$ for $l$-CNT in the range 0.3--2 THz, along with the obtained electron scattering rate $\gamma$ in figure \ref{fig:scat_rate}c.
	For temperatures below 500 K, the scattering rate from acoustic phonons is known to have a linear temperature dependence \cite{Zhou2005,Kane1998,Suzuura2002,Pennington2007a}, $\gamma_{e-ph}=\alpha T/d$, where $d$ is tube diameter and $\alpha$ is a constant. For our $l$-CNTs, the average diameter is 1.6 nm, giving $\alpha$ = 49 m/(Ks) from fitting $\gamma$ in the range 293--500 K (see dashed line in figure \ref{fig:scat_rate}c). This value of $\alpha$ corresponds to an electron mean free path of 109 nm at room temperature, which is close to that of graphite \cite{Herbert1997}.  The observed linear dependence $\gamma\propto T$ in this temperature range agrees well with this prediction, however the experimentally reported value of $\alpha$ = 12 m/(Ks) by Zhou et al. \cite{Zhou2005} is four times smaller than our obtained value for $\alpha$, indicating the electron-acoustic phonon scattering rate to be much smaller for their samples. The reason for this discrepancy is not currently known, but might be in part explained by the influence of contact-resistance in the measurements of \cite{Zhou2005}.
	
	In figure \ref{fig:scat_rate}d, the obtained electron--acoustic phonon term has been subtracted from the scattering rate $\gamma-\gamma_{e-ph}$. We see that at low temperatures ($<$ 200 K) there is a significant contribution from scattering mechanisms which do not originate from electron-acoustic phonon scattering or defect scattering, however at present time we do not know the origin of this scattering. We note that this temperature-dependent behaviour remained constant through multiple heating and cooling cycles, meaning it is reversible. Further investigation is required to determine the physical mechanisms responsible for this temperature dependence.
	
	\subsection{Percolation Effects}
\begin{figure}[tb]
		\centering
		\includegraphics[width=\linewidth]{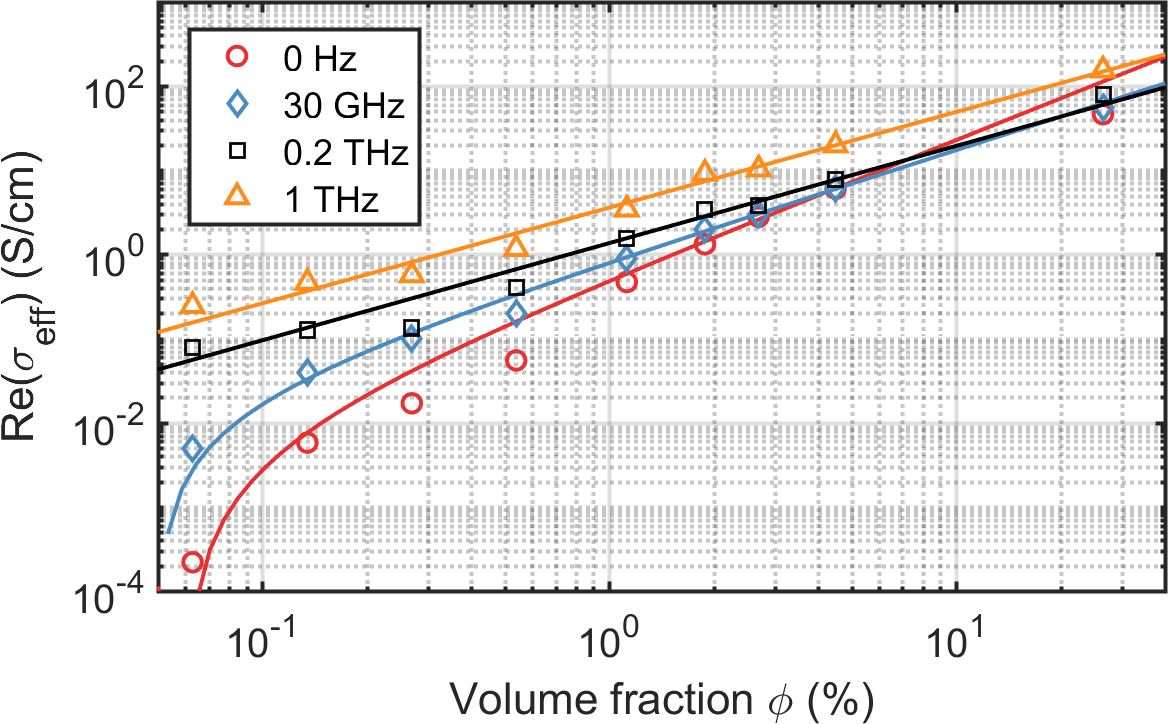}
		\caption{Effective conductivity Re$(\sigma_{eff})$ of $m$-CNT/INT-composites vs CNT volume fraction $\phi$ for frequencies 0Hz, 30GHz, 0.3THz and 1THz. Points are experimental data and lines are fitted power functions of the form $\sigma(\phi)=\sigma_0(\phi-\phi_0)^{\beta}$, with $\beta=1.6, 1.3, 1.2, 1.1$ and $\phi_0=0.06\%, 0.05\%, 0\%, 0\%$ for 0 Hz, 30 GHz, 0.3 THz and 1 THz, respectively.}
		\label{fig:xnewcondvsdens}
	\end{figure}

	While it is clear that temperature dependent scattering effects are sufficient to explain the conductivity temperature dependences of our CNT samples with tubes of varying length, this cannot by itself account for the huge variety of temperature dependences reported in the literature\cite{Morimoto2016,Thirunavukkuarasu2010,Borondics2006,Ugawa1999a,Zhang2013,Xu2007}. One cannot avoid the fact that in reality most CNT samples are percolating networks, with charge-carriers being able to travel between CNTs through electrical tunneling\cite{Gau2009,Ounaies2003}. The electrical properties of INT- and CNT/INT-composites have previously been studied from impedance measurements in the frequency range 0--$10^6$ Hz, showing significant dependence on the nanotube concentration \cite{Ksenevich2016,Lubineau2017}. Therefore the influence of percolation must be included. 
	In order to investigate this, the effective conductivity of $m$-CNT/INT-composites of varying nanotube volume fractions, $\phi$, were measured at 0 GHz, 30 GHz, 0.3 THz, and 1 THz, see figure \ref{fig:xnewcondvsdens}. The volume fraction was roughly estimated from the known CNT density $\rho_{CNT}$ of our film using the relation $\phi=\rho_{CNT}/\rho_{graphite}$, where $\rho_{graphite}=2.226$ g/cm$^3$ is the density of graphite. A significant result is the fact that our $m$-CNT/INT films show one order of magnitude higher conductivities compared with that reported for CNT-based composites of similar average volume fractions\cite{NingLi2006,Gong2013}. This is due to the CNTs being non-aggregated and well dispersed among the INTs, as demonstrated earlier in section \ref{sec:sampleprep}. Furthermore, let us note that the addition of fullerene-like WS$_2$ nanoparticles was recently shown to be efficient as an alternative route to produce well-dispersed CNT-based polymer nanocomposite \cite{Naffakh,Pascual}.
	
	The conductivity in figure \ref{fig:xnewcondvsdens} is shown to follow a power-law dependence $\sigma(\phi)=\sigma_0(\phi-\phi_0)^{\beta}$, where $\sigma_0$ is an arbitrary scaling factor, $\phi$ is the volume fraction of the CNT network, $\phi_0$ is the percolation threshold below which the sample effectively consists of isolated CNT bundles. Our fits of the data in figure \ref{fig:xnewcondvsdens} agrees with conductivity measurements of similar percolating nanotube networks showing a universal power-law scaling with frequency, which is common for systems with randomly distributed barriers \cite{Battisti2010,NingLi2006,Wen2013,Kim2003,Sandler2003}. From our fits we obtain $\beta=1.6, 1.3, 1.2, 1.1$ and $\phi_0=0.06\%, 0.05\%, 0\%, 0\%$ for 0 Hz, 30 GHz, 0.3 THz and 1 THz, respectively. The 0 Hz percolation threshold agrees well with similar studies on CNT-composite systems\cite{NingLi2006,Kim2003} and the difference in percolation thresholds and the decreasing $\beta$ for higher frequencies is expected due to an increasing sensitivity to the intrinsic conductivity of the nanotube\cite{Kim2003,Ounaies2003}.
	\begin{figure}[tb]
		\centering
		\includegraphics[width=1\linewidth]{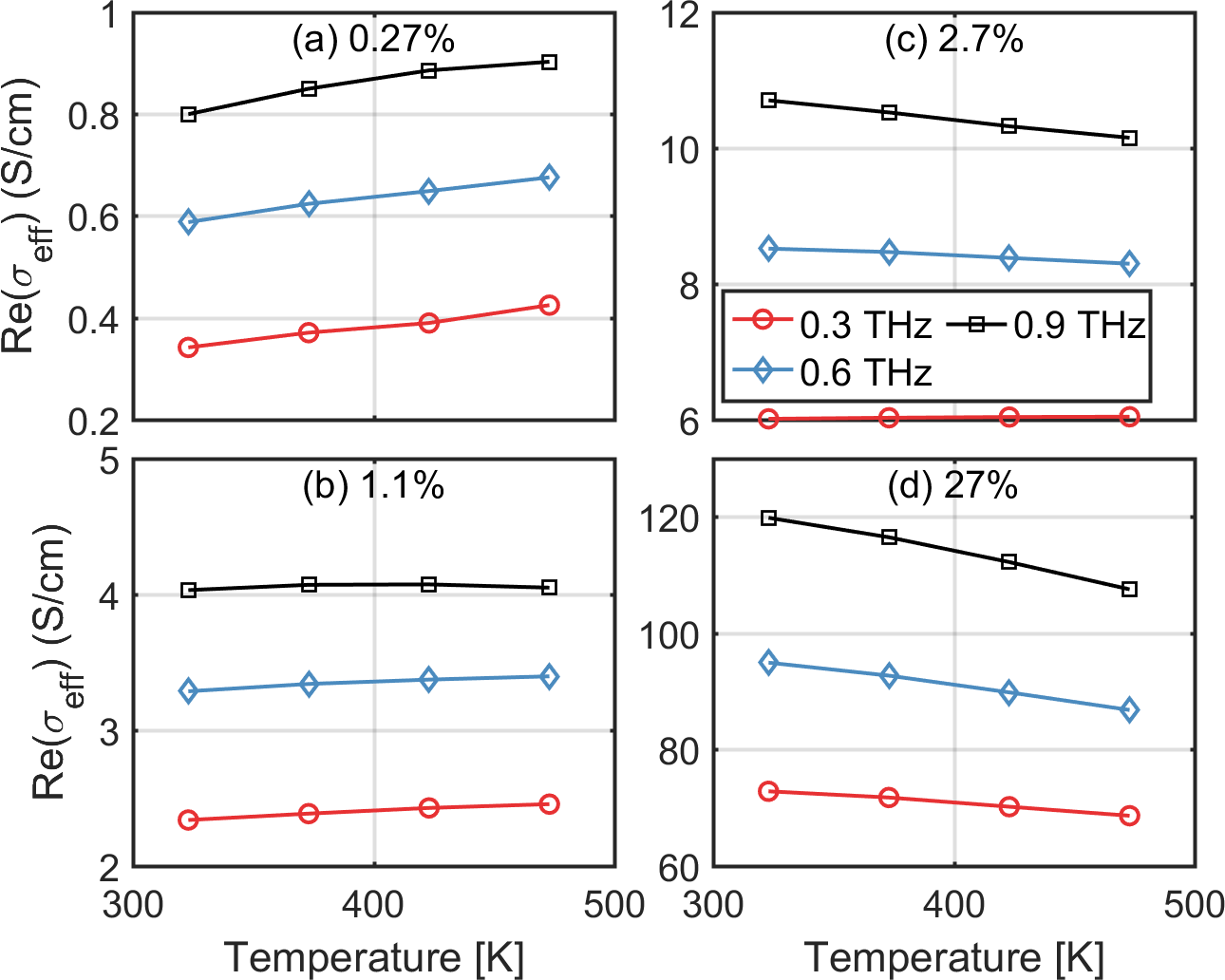}
		\caption{Effective conductivity Re$(\sigma_{eff})$ for $m$-CNT/INT-films versus temperature for volume fractions $\phi \in(0.27\%, 1.1\%, 2.7\%, 27\%)$.}
		\label{fig:cvst}
	\end{figure}		
	
	In addition to affecting the frequency dependence of the conductivity, percolation also influences the temperature dependence, which is illustrated in figure \ref{fig:cvst}, where ${\rm Re}(\sigma_{eff})$, obtained via THz-TDS, is plotted vs temperature for different CNT volume fractions $\phi$. It can be seen that as $\phi$ increases, the temperature dependence reverses. This appears, as in our model from section \ref{sec:theory}, because  the THz resonance shifts to lower frequencies for increasing $\phi$. This can again be explained by the finite-length effect, where the increasing CNT volume fraction results in an increasing number of contact points between nanotubes, thereby increasing the ``effective" nanotube length. Thus, our data for $m$-CNT shows (i) that percolation can strongly affect the temperature dependence for the effective conductivity in the THz range, (ii) the proportionality Re$(\sigma_{eff})\propto\phi$ is true in the range 0.2--1 THz giving us a method to determine CNT volume fraction from THz-TDS measurements \cite{Peters2012}. (iii) The volume fraction (or density) of the CNTs can strongly influence effective tube length and thus the location of the effective THz conductivity peak. Finally, since INTs demonstrate low conductivity below 3 THz and high temperature stability, they can be used as an ``inert" material for the fabrication of hybrid films with randomly dispersed elongated conductive nanoparticles. Here we have demonstrated for the first time that such hybrid films are a great tool for studying the percolation effect in a wide range of temperatures, frequencies and concentrations.
	
	\section{Conclusion}
	In summary, free-standing films comprising single- walled CNTs of various lengths, and CNT/WS$_2$ INT composites of varying CNT densities on quartz, were prepared and their effective conductivities and permittivities measured in the ranges (i) 0.2--1000 THz for 300--530 K, and (ii) 0.3--2 THz for 10--530 K. We observe a peak in the THz conductivity spectra of the CNT films which is shown to be plasmonic in nature, in agreement with recent literature, and  shifting in frequency with the CNT length as predicted by our model of the localized plasmon resonance. Drude fits of the THz conductivity spectra for long ($L\approx$ 10 $\mu$m) CNTs show a linear dependence of the electron scattering rate $\gamma\propto$ T for 300--500 K, revealing electron-acoustic phonon scattering as the main scattering mechanism in the individual CNTs in this range \cite{Zhou2005,Kane1998}. Due to the plasmonic nature of the response of shorter tubes ($L\le$ 1 $\mu$m), we observe an unexpected variation with temperature, where the effective conductivity of the films in the THz frequency range is shown to increase with temperature, while we observe a decrease in temperature for longer tubes. This temperature-induced effect arises indirectly from the variation of the electron scattering rate, which determines plasmon losses and thus broadening of the conductivity resonance. Moreover, by studying composites of varying CNT densities, we show that the change in conductivity with temperature depends not only on tube length, but also varies with tube density: as the temperature increases, we observe an effective conductivity in the range 0.3--1 THz that increases/decreases for low/high density samples. This effect accounts for the apparent discrepancies in literature regarding temperature dependent conductivity of CNT composites \cite{Morimoto2016,Thirunavukkuarasu2010,Borondics2006,Ugawa1999a,Zhang2013,Xu2007}, and occurs due to the density dependence of the effective length of conducting pathways in the composite films, which again leads to a shift and temperature dependent broadening of the THz conductivity resonance. Our CNT/INT composite films were characterised using an SEM, showing well-dispersed and non-aggregated CNTs, and the composites demonstrate approximately an order of magnitude higher effective conductivities compared with CNT-composites with similar average volume fractions \cite{NingLi2006,Gong2013}. Due to their low conductivity below 3 THz and their high temperature stability, INTs show great potential as an ``inert" material for fabrication of hybrid films with randomly dispersed and elongated conductive nanoparticles, and as a tool for studying the effects of percolation.  
	
	\ack{}
	This research was partially supported by the Belarusian Republican Foundation for Fundamental Research (BRFFR) under project F15CO-016, the European Union’s Seventh Framework Programme (FP7) for research, technological development and demonstration under	projects FP7-612285 CANTOR, PIRSES-GA-2012-318617, 607521 NOTEDEV, and by the H2020-MSCA-RISE-2014 project 644076 CoExAN. Russian Science Foundation is greatly acknowledged for financial support (agreement No 17-19-01787). RT acknowledges the support of the H. Perlman foundation and the Irving and Azelle Waltcher Foundations in honour of Prof. M. Levy.
	
	\section*{References}
	\bibliographystyle{iopart-num} % or try abbrvnat or unsrtnat
	\bibliography{literature}

\providecommand{\newblock}{}
\begin{thebibliography}{10}
\expandafter\ifx\csname url\endcsname\relax
  \def\url#1{{\tt #1}}\fi
\expandafter\ifx\csname urlprefix\endcsname\relax\def\urlprefix{URL }\fi
\providecommand{\eprint}[2][]{\url{#2}}

\bibitem{Reich2008}
Reich S, Thomsen C and Maultzsch J 2008 {\em Carbon Nanotubes: Basic Concepts
  and Physical Properties\/} (Wiley) ISBN 9783527618057
  \urlprefix\url{https://books.google.by/books?id=nlRWOE6C0AsC}

\bibitem{Portnoi2008}
Portnoi M, Kibis O and {Rosenau da Costa} M 2008 {\em Superlattices and
  Microstructures\/} {\bf 43} 399--407 ISSN 07496036
  \urlprefix\url{http://www.sciencedirect.com/science/article/pii/S0749603607002108}

\bibitem{He2014}
He X, Fujimura N, Lloyd J~M, Erickson K~J, Talin A~A, Zhang Q, Gao W, Jiang Q,
  Kawano Y, Hauge R~H, L{\'{e}}onard F and Kono J 2014 {\em Nano letters\/}
  {\bf 14} 3953--8 ISSN 1530-6992
  \urlprefix\url{http://dx.doi.org/10.1021/nl5012678}

\bibitem{Titova2015}
Titova L~V, Pint C~L, Zhang Q, Hauge R~H, Kono J and Hegmann F~A 2015 {\em Nano
  letters\/} ISSN 1530-6992
  \urlprefix\url{http://dx.doi.org/10.1021/acs.nanolett.5b00494}

\bibitem{Zubair2016}
Zubair A, Tsentalovich D~E, Young C~C, Heimbeck M~S, Everitt H~O, Pasquali M
  and Kono J 2016 {\em Applied Physics Letters\/} {\bf 108} 141107 ISSN
  0003-6951 \urlprefix\url{http://aip.scitation.org/doi/10.1063/1.4945708}

\bibitem{Brosseau2012}
Brosseau F and Brosseau C 2012 {\em Journal of Applied Physics\/} {\bf 111}
  061301 (\textit{Preprint} \eprint{http://dx.doi.org/10.1063/1.3688435})
  \urlprefix\url{http://dx.doi.org/10.1063/1.3688435}

\bibitem{Hartmann2014}
Hartmann R~R, Kono J and Portnoi M~E 2014 {\em Nanotechnology\/} {\bf 25}
  322001 ISSN 1361-6528
  \urlprefix\url{http://www.ncbi.nlm.nih.gov/pubmed/25051014}

\bibitem{Shuba2017}
Shuba M~V, Paddubskaya A~G, Kuzhir P~P, Maksimenko S~A, Flahaut E, Fierro V,
  Celzard A and Valusis G 2017 {\em Journal of Physics D: Applied Physics\/}
  {\bf 50} 08LT01
  \urlprefix\url{http://stacks.iop.org/0022-3727/50/i=8/a=08LT01}

\bibitem{Bommeli1997}
Bommeli F, Degiorgi L, Wachter P, Bacsa W, de~Heer W and Forro L 1997 {\em
  Synthetic Metals\/} {\bf 86} 2307 -- 2308 ISSN 0379-6779 proceedings of the
  International Conference on Science and Technology of Synthetic Metals
  \urlprefix\url{http://www.sciencedirect.com/science/article/pii/S037967799781137X}

\bibitem{Kane1997}
Kane C~L and Mele E~J 1997 {\em Phys. Rev. Lett.\/} {\bf 78}(10) 1932--1935
  \urlprefix\url{https://link.aps.org/doi/10.1103/PhysRevLett.78.1932}

\bibitem{Kampfrath2008}
Kampfrath T, von Volkmann K, Aguirre C~M, Desjardins P, Martel R, Krenz M,
  Frischkorn C, Wolf M and Perfetti L 2008 {\em Phys. Rev. Lett.\/} {\bf 101}
  267403 ISSN 0031-9007
  \urlprefix\url{http://link.aps.org/doi/10.1103/PhysRevLett.101.267403}

\bibitem{Ugawa1999}
Ugawa A, Rinzler A~G and Tanner D~B 1999 {\em Phys. Rev. B\/} {\bf 60}
  R11305--R11308 ISSN 1098-0121
  \urlprefix\url{http://link.aps.org/doi/10.1103/PhysRevB.60.R11305}

\bibitem{Akima2006}
Akima N, Iwasa Y, Brown S, Barbour A, Cao J, Musfeldt J, Matsui H, Toyota N,
  Shiraishi M, Shimoda H and Zhou O 2006 {\em Advanced Materials\/} {\bf 18}
  1166--1169 ISSN 1521-4095
  \urlprefix\url{http://dx.doi.org/10.1002/adma.200502505}

\bibitem{Slepyan2006}
Slepyan G~Y, Shuba M~V, Maksimenko S~A and Lakhtakia A 2006 {\em Phys. Rev.
  B\/} {\bf 73} 195416 ISSN 1098-0121
  \urlprefix\url{http://link.aps.org/doi/10.1103/PhysRevB.73.195416}

\bibitem{Slepyan2010a}
Slepyan G~Y, Shuba M~V, Maksimenko S~A, Thomsen C and Lakhtakia A 2010 {\em
  Phys. Rev. B\/} {\bf 81} 205423 ISSN 1098-0121
  \urlprefix\url{http://link.aps.org/doi/10.1103/PhysRevB.81.205423}

\bibitem{Shuba2012}
Shuba M~V, Paddubskaya A~G, Plyushch A~O, Kuzhir P~P, Slepyan G~Y, Maksimenko
  S~A, Ksenevich V~K, Buka P, Seliuta D, Kasalynas I, Macutkevic J, Valusis G,
  Thomsen C and Lakhtakia A 2012 {\em Phys. Rev. B\/} {\bf 85} 14--17 ISSN
  1098-0121 \urlprefix\url{http://link.aps.org/doi/10.1103/PhysRevB.85.165435}

\bibitem{Zhang2013}
Zhang Q, H{\'{a}}roz E~H, Jin Z, Ren L, Wang X, Arvidson R~S, L{\"{u}}ttge A
  and Kono J 2013 {\em Nano letters\/} {\bf 13} 5991--6 ISSN 1530-6992
  (\textit{Preprint} \eprint{arXiv:1312.4067v1})
  \urlprefix\url{http://pubs.acs.org/doi/abs/10.1021/nl403175g}

\bibitem{Kim2003}
Kim B, Lee J and Yu I 2003 {\em Journal of Applied Physics\/} {\bf 94}
  6724--6728 ISSN 00218979

\bibitem{Xu2009}
Xu X, Chuang K, Nicholas R~J, Johnston M~B and Herz L~M 2009 {\em Journal of
  Physical Chemistry C\/} {\bf 113} 18106--18109 ISSN 19327447
  \urlprefix\url{http://pubs.acs.org/doi/abs/10.1021/jp907195t}

\bibitem{Bauhofer2009}
Bauhofer W and Kovacs J~Z 2009 {\em Composites Science and Technology\/} {\bf
  69} 1486--1498 ISSN 02663538

\bibitem{Wen2013}
Wen B, Cao M~S, Hou Z~L, Song W~L, Zhang L, Lu M~M, Jin H~B, Fang X~Y, Wang W~Z
  and Yuan J 2013 {\em Carbon\/} {\bf 65} 124--139 ISSN 00086223

\bibitem{Borondics2006}
Borondics F, Kamar{\'{a}}s K, Nikolou M, Tanner D~B, Chen Z~H and Rinzler A~G
  2006 {\em Physical Review B\/} {\bf 74} 45431 ISSN 1098-0121
  (\textit{Preprint} \eprint{0512667})
  \urlprefix\url{https://link.aps.org/doi/10.1103/PhysRevB.74.045431}

\bibitem{Thirunavukkuarasu2010}
Thirunavukkuarasu K, Hennrich F, Kamar{\'{a}}s K and Kuntscher C~A 2010 {\em
  Phys. Rev. B\/} {\bf 81} 1--12 ISSN 1098-0121
  \urlprefix\url{http://link.aps.org/doi/10.1103/PhysRevB.81.045424}

\bibitem{Morimoto2016}
Morimoto T, Ichida M, Ikemoto Y and Okazaki T 2016 {\em Phys. Rev. B\/} {\bf
  93} 195409 ISSN 2469-9950
  \urlprefix\url{http://link.aps.org/doi/10.1103/PhysRevB.93.195409}

\bibitem{Xu2007}
Xu H, Anlage S~M, Hu L and Gruner G 2007 {\em Applied Physics Letters\/} {\bf
  90} 3--5 ISSN 00036951

\bibitem{Shuba2016}
Shuba M~V, Paddubskaya A~G, Kuzhir P~P, Maksimenko S~A, Valusis G, Poklonski
  N~A, Bellucci S, Kenanakis G and Kafesaki M 2016 {\em Journal of Applied
  Physics\/} {\bf 119} 104303 ISSN 0021-8979
  \urlprefix\url{http://scitation.aip.org/content/aip/journal/jap/119/10/10.1063/1.4943499}

\bibitem{Ugawa1999a}
Ugawa A, Rinzler A~G and Tanner D~B 1999 {\em Phys. Rev. B\/} {\bf 60}
  R11305--R11308 ISSN 0163-1829
  \urlprefix\url{http://link.aps.org/doi/10.1103/PhysRevB.60.R11305}

\bibitem{Kampfrath2007a}
Kampfrath T, Perfetti L, von Volkmann K, Aguirre C~M, Desjardins P, Martel R,
  Frischkorn C and Wolf M 2007 {\em Physica Status Solidi (B) Basic Research\/}
  {\bf 244} 3950--3954 ISSN 03701972
  \urlprefix\url{http://doi.wiley.com/10.1002/pssb.200776112}

\bibitem{Nishimura2007}
Nishimura H, Minami N and Shimano R 2007 {\em Applied Physics Letters\/} {\bf
  91} 2005--2008 ISSN 00036951
  \urlprefix\url{http://scitation.aip.org/content/aip/journal/apl/91/1/10.1063/1.2753747}

\bibitem{Nemilentsau2010}
Nemilentsau A~M, Shuba M~V, Slepyan G~Y, Kuzhir P~P, Maksimenko S~A, D'yachkov
  P~N and Lakhtakia A 2010 {\em Phys. Rev. B\/} {\bf 82} 235424 ISSN 1098-0121
  \urlprefix\url{https://link.aps.org/doi/10.1103/PhysRevB.82.235424}

\bibitem{Slepyan1999a}
Slepyan G~Y, Maksimenko S~A, Lakhtakia A, Yevtushenko O and Gusakov A~V 1999
  {\em Phys. Rev. B\/} {\bf 60} 17136--17149 ISSN 0163-1829
  \urlprefix\url{http://link.aps.org/doi/10.1103/PhysRevB.60.17136}

\bibitem{Zhou2005a}
Zhou X, Park J~Y, Huang S, Liu J and McEuen P~L 2005 {\em Phys. Rev. Lett.\/}
  {\bf 95} 146805
  \urlprefix\url{https://link.aps.org/doi/10.1103/PhysRevLett.95.146805}

\bibitem{Kane1998}
Kane C~L, Mele E~J, Lee R~S, Fischer J~E, Petit P, Dai H, Thess A, Smalley R~E,
  Verschueren A~R~M, Tans S~J and Dekker C 1998 {\em Europhysics Letters
  (EPL)\/} {\bf 41} 683--688 ISSN 0295-5075
  \urlprefix\url{http://stacks.iop.org/0295-5075/41/i=6/a=683?key=crossref.a6c8c0050f7153d716f148b12e0d2b05}

\bibitem{Zhou2005}
Zhou X, Park J~Y, Huang S, Liu J and McEuen P~L 2005 {\em Phys. Rev. Lett.\/}
  {\bf 95} 146805 ISSN 0031-9007
  \urlprefix\url{https://link.aps.org/doi/10.1103/PhysRevLett.95.146805}

\bibitem{Shuba2007}
Shuba M~V, Maksimenko S~A and Lakhtakia A 2007 {\em Phys. Rev. B\/} {\bf 76}
  155407 ISSN 1098-0121
  \urlprefix\url{https://link.aps.org/doi/10.1103/PhysRevB.76.155407}

\bibitem{Nasibulin2011}
Moisala A, Nasibulin A~G, Brown D~P, Jiang H, Khriachtchev L and Kauppinen E~I
  2006 {\em Chemical Engineering Science\/} {\bf 61} 4393 -- 4402 ISSN
  0009-2509 the John Bridgwater Symposium: "Shaping the Future of Chemical
  Engineering"
  \urlprefix\url{http://www.sciencedirect.com/science/article/pii/S0009250906001278}

\bibitem{Tian2011}
Tian Y, Timmermans M~Y, Kivist{\"o} S, Nasibulin A~G, Zhu Z, Jiang H,
  Okhotnikov O~G and Kauppinen E~I 2011 {\em Nano Research\/} {\bf 4} 807 ISSN
  1998-0000 \urlprefix\url{https://doi.org/10.1007/s12274-011-0137-6}

\bibitem{Hennrich2002}
Hennrich F, Lebedkin S, Malik S, Tracy J, Barczewski M, Rosner H and Kappes M
  2002 {\em Phys. Chem. Chem. Phys.\/} {\bf 4}(11) 2273--2277
  \urlprefix\url{http://dx.doi.org/10.1039/B201570F}

\bibitem{Hennrich2007}
Hennrich F, Krupke R, Arnold K, Rojas~Stütz J~A, Lebedkin S, Koch T, Schimmel
  T and Kappes M~M 2007 {\em The Journal of Physical Chemistry B\/} {\bf 111}
  1932--1937 pMID: 17274643 (\textit{Preprint}
  \eprint{http://dx.doi.org/10.1021/jp065262n})
  \urlprefix\url{http://dx.doi.org/10.1021/jp065262n}

\bibitem{Shuba2012N}
Shuba M~V, Paddubskaya A~G, Kuzhir P~P, Maksimenko S~A, Ksenevich V~K, Niaura
  G, Seliuta D, Kasalynas I and Valusis G 2012 {\em Nanotechnology\/} {\bf 23}
  495714 \urlprefix\url{http://stacks.iop.org/0957-4484/23/i=49/a=495714}

\bibitem{Coleman1998}
Coleman J~N, Curran S, Dalton A~B, Davey A~P, McCarthy B, Blau W and Barklie
  R~C 1998 {\em Phys. Rev. B\/} {\bf 58} R7492--R7495 ISSN 0163-1829
  \urlprefix\url{http://link.aps.org/doi/10.1103/PhysRevB.58.R7492}

\bibitem{Zak2009}
Zak A, Sallan-Ecker L, Margolin A, Genut M and Tenne R 2009 {\em Nano\/} {\bf
  04} 91--98 ISSN 1793-2920
  \urlprefix\url{http://www.worldscientific.com/doi/abs/10.1142/S1793292009001551}

\bibitem{Ulbricht2011}
Ulbricht R, Hendry E, Shan J, Heinz T~F and Bonn M 2011 {\em Reviews of Modern
  Physics\/} {\bf 83} 543--586 ISSN 0034-6861
  \urlprefix\url{http://link.aps.org/doi/10.1103/RevModPhys.83.543}

\bibitem{GHz}
Standard test method for measuring relative complex permittivity and relative
  magnetic permeability of solid materials at microwave frequencies, ASTM
  D5568-08, 2009.

\bibitem{Chung2007}
Chung B~K 2007 {\em Progress In Electromagnetics Research\/} {\bf 75} 239
  \urlprefix\url{http://onlinewww.jpier.org/pier/pier.php?paper=07052801}

\bibitem{Paddubskaya2016}
Paddubskaya A, Valynets N, Kuzhir P, Batrakov K, Maksimenko S, Kotsilkova R,
  Velichkova H, Petrova I, Bir{\'{o}} I, Kert{\'{e}}sz K, M{\'{a}}rk G~I,
  Horv{\'{a}}th Z~E and Bir{\'{o}} L~P 2016 {\em Journal of Applied Physics\/}
  {\bf 119} 135102 ISSN 0021-8979
  \urlprefix\url{http://aip.scitation.org/doi/10.1063/1.4945576}

\bibitem{PhysRevB.60.R11305}
Ugawa A, Rinzler A~G and Tanner D~B 1999 {\em Phys. Rev. B\/} {\bf 60}(16)
  R11305--R11308
  \urlprefix\url{https://link.aps.org/doi/10.1103/PhysRevB.60.R11305}

\bibitem{Suzuura2002}
Suzuura H and Ando T 2002 {\em Phys. Rev. B\/} {\bf 65} 235412 ISSN 0163-1829
  \urlprefix\url{http://link.aps.org/doi/10.1103/PhysRevB.65.235412}

\bibitem{Pennington2007a}
Pennington G, Goldsman N, Akturk A and Wickenden A~E 2007 {\em Applied Physics
  Letters\/} {\bf 90} ISSN 00036951 (\textit{Preprint} \eprint{0610777})
  \urlprefix\url{http://arxiv.org/abs/cond-mat/0610777
  http://dx.doi.org/10.1063/1.2437127}

\bibitem{Herbert1997}
Philipp H~R 1977 {\em Phys. Rev. B\/} {\bf 16}(6) 2896--2900
  \urlprefix\url{https://link.aps.org/doi/10.1103/PhysRevB.16.2896}

\bibitem{Gau2009}
Gau C, Kuo C~Y and Ko H~S 2009 {\em Nanotechnology\/} {\bf 20} 395705 ISSN
  0957-4484
  \urlprefix\url{http://stacks.iop.org/0957-4484/20/i=39/a=395705?key=crossref.06b3c84620bbb966c37e88dd375aee15}

\bibitem{Ounaies2003}
Ounaies Z, Park C, Wise K~E, Siochi E~J and Harrison J~S 2003 {\em Composites
  Science and Technology\/} {\bf 63} 1637--1646 ISSN 02663538

\bibitem{Ksenevich2016}
Ksenevich V, Gorbachuk N, Viet H, Shuba M, Kuzhir P, Maksimenko S, Paddubskaya
  A, Valusis G, Wieck A, Zak A and Tenne R 2016 {\em Nanosystems: Physics,
  Chemistry, Mathematics\/} {\bf 7} 37--43 ISSN 22208054
  \urlprefix\url{http://nanojournal.ifmo.ru/en/articles-2/volume7/7-1/paper03/}

\bibitem{Lubineau2017}
Lubineau G, Mora A, Han F, Odeh I and Yaldiz R 2017 {\em Computational
  Materials Science\/} {\bf 130} 21--38 ISSN 09270256

\bibitem{NingLi2006}
Li N, Huang Y, Du F, He X, Lin X, Gao H, Ma Y, Li F, Chen Y and Eklund P~C 2006
  {\em Nano Letters\/} {\bf 6} 1141--1145 ISSN 15306984
  \urlprefix\url{http://pubs.acs.org/doi/abs/10.1021/nl0602589}

\bibitem{Gong2013}
Gong S, Zhu Z~H and Haddad E~I 2013 {\em Journal of Applied Physics\/} {\bf
  114} 074303 ISSN 00218979
  \urlprefix\url{http://scitation.aip.org/content/aip/journal/jap/114/7/10.1063/1.4818478}

\bibitem{Naffakh}
Naffakh M, Diez-Pascual A~M and Gomez-Fatou M~A 2011 {\em J. Mater. Chem.\/}
  {\bf 21}(20) 7425--7433 \urlprefix\url{http://dx.doi.org/10.1039/C1JM10441A}

\bibitem{Pascual}
Díez-Pascual A~M, Naffakh M, Marco C and Ellis G 2012 {\em The Journal of
  Physical Chemistry B\/} {\bf 116} 7959--7969 pMID: 22697425
  (\textit{Preprint} \eprint{http://dx.doi.org/10.1021/jp3035314})
  \urlprefix\url{http://dx.doi.org/10.1021/jp3035314}

\bibitem{Battisti2010}
Battisti A, Skordos A~A and Partridge I~K 2010 {\em Composites Science and
  Technology\/} {\bf 70} 633--637 ISSN 02663538
  \urlprefix\url{http://linkinghub.elsevier.com/retrieve/pii/S0266353809004527}

\bibitem{Sandler2003}
Sandler J, Kirk J, Kinloch I, Shaffer M and Windle A 2003 {\em Polymer\/} {\bf
  44} 5893--5899 ISSN 00323861

\bibitem{Peters2012}
Peters O, Busch S~F, Fischer B~M and Koch M 2012 {\em Journal of Infrared,
  Millimeter, and Terahertz Waves\/} {\bf 33} 1221--1226 ISSN 1866-6906
  \urlprefix\url{https://doi.org/10.1007/s10762-012-9935-5}

\end{thebibliography}


\providecommand{\newblock}{}
\begin{thebibliography}{1}
\expandafter\ifx\csname url\endcsname\relax
  \def\url#1{{\tt #1}}\fi
\expandafter\ifx\csname urlprefix\endcsname\relax\def\urlprefix{URL }\fi
\providecommand{\eprint}[2][]{\url{#2}}

\bibitem{Zhang2013}
Zhang Q, H{\'{a}}roz E~H, Jin Z, Ren L, Wang X, Arvidson R~S, L{\"{u}}ttge A
  and Kono J 2013 {\em Nano letters\/} {\bf 13} 5991--6 ISSN 1530-6992
  (\textit{Preprint} \eprint{arXiv:1312.4067v1})
  \urlprefix\url{http://pubs.acs.org/doi/abs/10.1021/nl403175g}

\bibitem{Shuba2016}
Shuba M~V, Paddubskaya A~G, Kuzhir P~P, Maksimenko S~A, Valusis G, Poklonski
  N~A, Bellucci S, Kenanakis G and Kafesaki M 2016 {\em Journal of Applied
  Physics\/} {\bf 119} 104303 ISSN 0021-8979
  \urlprefix\url{http://scitation.aip.org/content/aip/journal/jap/119/10/10.1063/1.4943499}

\bibitem{Miyata2011}
Miyata Y, Mizuno K and Kataura H 2011 {\em Journal of Nanomaterials\/} {\bf
  2011} 1--7 ISSN 1687-4110
  \urlprefix\url{http://www.hindawi.com/journals/jnm/2011/786763/}

\bibitem{Jorio2003}
Jorio A, Pimenta M~A, Filho A~G~S, Saito R, Dresselhaus G and Dresselhaus M~S
  2003 {\em New Journal of Physics\/} {\bf 5} 139--139 ISSN 1367-2630
  \urlprefix\url{http://stacks.iop.org/1367-2630/5/i=1/a=139}

\bibitem{Ulbricht2011}
Ulbricht R, Hendry E, Shan J, Heinz T~F and Bonn M 2011 {\em Reviews of Modern
  Physics\/} {\bf 83} 543--586 ISSN 0034-6861
  \urlprefix\url{http://link.aps.org/doi/10.1103/RevModPhys.83.543}

\bibitem{GHz}
Standard test method for measuring relative complex permittivity and relative
  magnetic permeability of solid materials at microwave frequencies, ASTM
  D5568-08, 2009.

\bibitem{Chung2007}
Chung B~K 2007 {\em Progress In Electromagnetics Research\/} {\bf 75} 239
  \urlprefix\url{http://onlinewww.jpier.org/pier/pier.php?paper=07052801}

\end{thebibliography}
	
\end{document}

\title[Supplementary Material]{Supplementary Material for ``Influence of nanotube length and density on the plasmonic terahertz response of single-walled carbon nanotubes"}
	\author{P Karlsen$^1$, M V Shuba$^2$, C Beckerleg$^1$, D I Yuko$^2$, P P Kuzhir$^2$, S A Maksimenko$^2$, V Ksenevich$^3$, Ho Viet$^3$, A G Nasibulin$^{4,5}$, R Tenne$^6$, and E Hendry$^1$}
	\address{$^1$ School of Physics, University of Exeter, Stocker Road, EX4 4QL, United Kingdom}
	\address{$^2$ Institute for Nuclear Problems, Belarus State University, Bobruiskaya 11, 220050 Minsk, Belarus}
	\address{$^3$ Department of Physics, Belarus State University, Nezalezhnastsi Avenue 4, 220030 Minsk, Belarus}
	\address{$^4$ Skolkovo Institute of Science and Technology, Skolkovo Innovation Center, Building 3, Moscow 143026, Russia}
	\address{$^5$ Department of Applied Physics, Aalto University, School of Science, P.O. Box 15100, FI-00076 Espoo, Finland}
	\address{$^6$ Department of Materials and Interfaces, Weizmann Institute of Science, Rehovot 76100, Israel}
	\eads{\mailto{peterkarlsen88@gmail.com}, \mailto{mikhail.shuba@gmail.com}}
	\vspace{2pc}
	\noindent{\it Keywords}: Carbon Nanotubes, Terahertz, Dielectric Properties, Percolation, Temperature, Plasmon, Tungsten Disulfide 
	%\maketitle
	\ioptwocol
	\renewcommand{\thefigure}{S\arabic{figure}}
	\renewcommand{\thesection}{S\arabic{section}}
	\setcounter{figure}{0}
	
	\section{Comparison of bundled and unbundled CNTs}\label{sec:S_unbundledvsbundled}
\begin{figure}
		\centering
		\includegraphics[width=\linewidth]{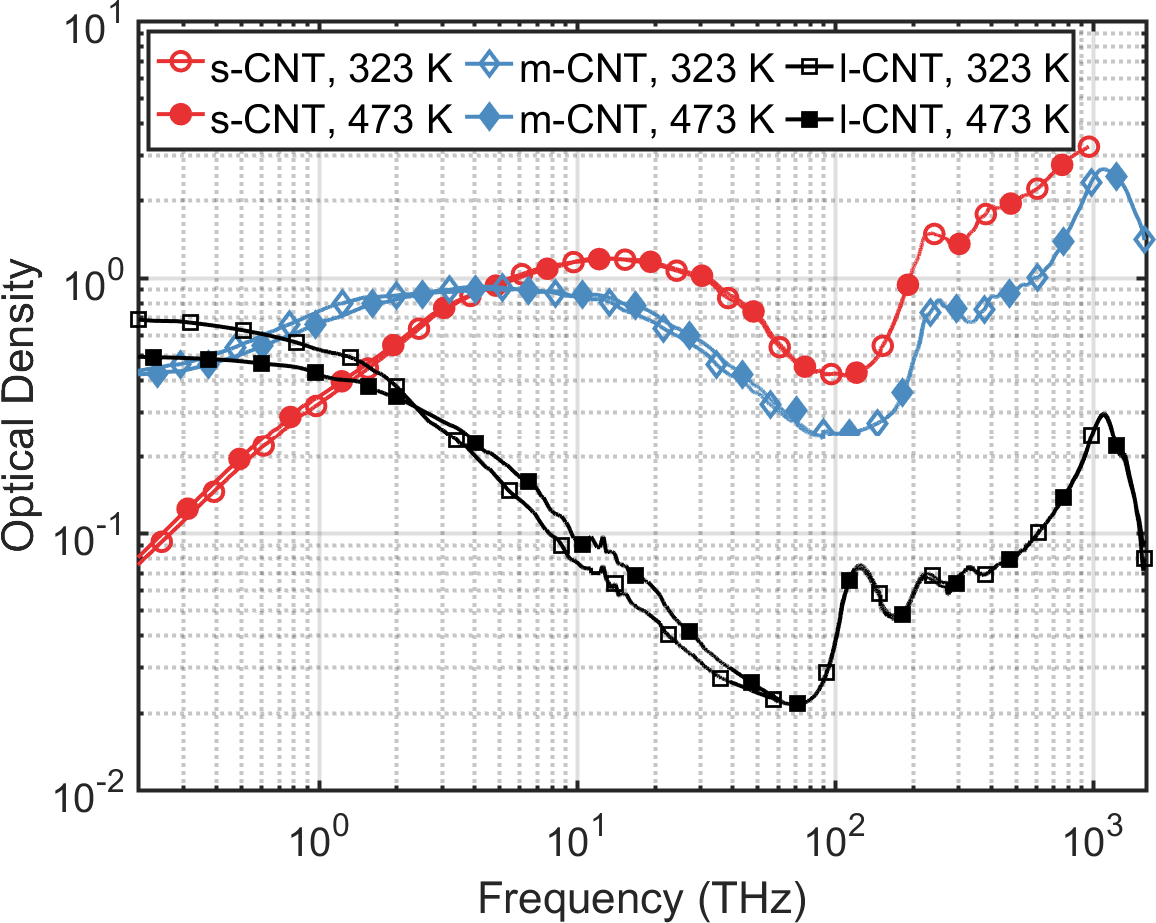}
		\caption{Frequency dependence of the optical density at 323 K and 473 K for $s$-, $m$-, and $l$-CNT films.}
		\label{fig:od}
	\end{figure}

\begin{figure}
		\centering
		\includegraphics[width=\linewidth]{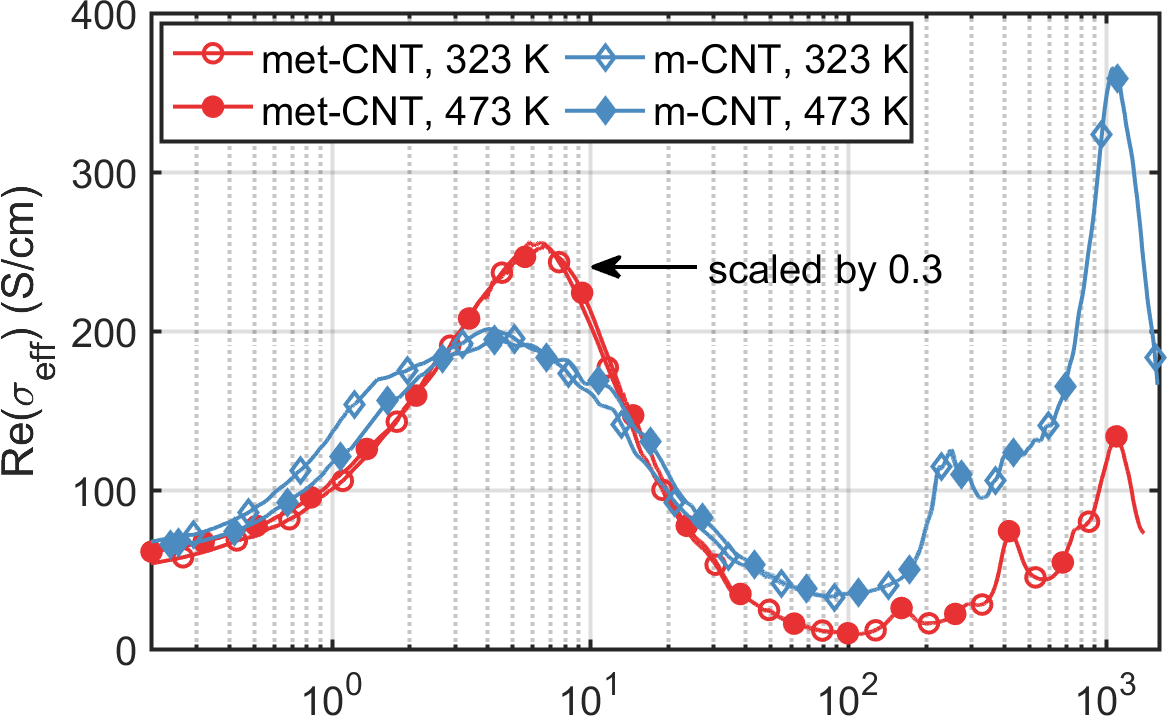}
		\caption{Frequency dependence of the real part of the effective conductivity Re$(\sigma_{eff})$, obtained from broadband optical density measurements at 323 K and 473 K for $met$-,and $m$-CNT films. Note that the data for the $met$-CNTs have been scaled by 0.3.}
		\label{fig:odcondcompare}
	\end{figure}

	As described in section 4 in the main text, the optical density spectra of our CNT films were obtained with normally incident light in the ranges 0.1--2 THz, 2--270 THz, and 270--1000 THz  using a terahertz time-domain spectrometer (EKSPLA, Lithuania), a  Fourier-transform infrared spectrometer Vertex 70 (Bruker), and an RV2201 spectrophotometer (ZAO SOLAR, Belarus), respectively. Figure \ref{fig:od} shows the optical density spectra for $s$-, $m$- and $l$-CNT films at 323 and 473 K, which we use to calculate the effective conductivity and permittivity in figure 4 using the Kramer-Kronig relations \cite{Zhang2013}. As shown in figure 4, we observe a shift of the THz peak to lower frequencies with increasing length, leading to a reversal of the temperature dependence for $s$-CNT and $l$-CNT around 1 THz. Note the slight blueshift of the THz-peak for $m$-CNTs with increasing temperature. The shift arises for bundled tubes  due to the temperature dependence of the conductivity of some fration of doped semiconducting CNTs with a Fermi level located close to the top of the valence band  \cite{Shuba2016}.   We demonstrate this shift theoretically and experimentally in Figures 2 and 4, respectively, however to further verify this, unbundled CNT-films ($met$-CNT) containing 95\% metallic CNTs (\href{http://www.nanointegris.com/metallic}{IsoNanotubes-M, purchased from NanoIntegris}) with average length of $L =$ 0.5 $\mu{}m$ were prepared using the same method as the $m$-CNT films, as described in section 3. We compare the effective conductivity Re($\sigma_{eff}$) of the bundled $m$-CNT and unbundled $met$-CNT films in figure \ref{fig:odcondcompare}, where we observe no blueshift of the THz peak for $met$-CNTs, as opposed to $m$-CNTs, which further demonstrates that the blueshift only occurs for bundled CNTs. Note that the $met$-CNT film shows an order of magnitude higher conductivity than the $m$-CNT film in the THz region, which is mainly due to the higher content of metallic CNTs in the former. This is in keeping with previous measurements of the THz conductivity for metallic and semiconducting CNT films \cite{Zhang2013}, where the former shows 2-3 times higher conductivity at the THz peak frequency.
	
	\section{Characterization of sample purity and crystalline quality}\label{sec:S_sample_quality}
	We investigate crystalline quality and purity of our CNT films and CNT/INT composites using a combination of Scanning Electron Microscopy (SEM) and Raman spectroscopy. SEM was used to check the purity of our samples, specifically if the CNTs are well-dispersed and non-aggregated, and if there are any carbonaceous impurities present, since these can effect the Raman intensity of the CNT samples \cite{Miyata2011}. Figure \ref{fig:SEM-CNT} shows SEM images of the m-CNT/INT and met-CNT/INT composites with varying volume fractions (see section \ref{sec:S_unbundledvsbundled} for details of $met$-CNTs). For both types of composites and for all volume fractions, the samples show successful dispersion and non-aggregation of the CNTs among the INTs, since the nanotubes are dispersed evenly in the sample and no dense particles compromising ten or more nanotubes, or any carbonaceous impurities, can be observed, indicating relatively high purity of our samples. Furthermore, we see that as the volume fraction of CNTs increases, the density of the CNT network increases evenly as well, thereby demonstrating the potential of our preparation-method for achieving high quality CNT/INT composites with a tunable, well-dispersed and non-aggregated CNT network. 
	
\begin{figure}[htb]
		\centering
		\includegraphics[width=\linewidth]{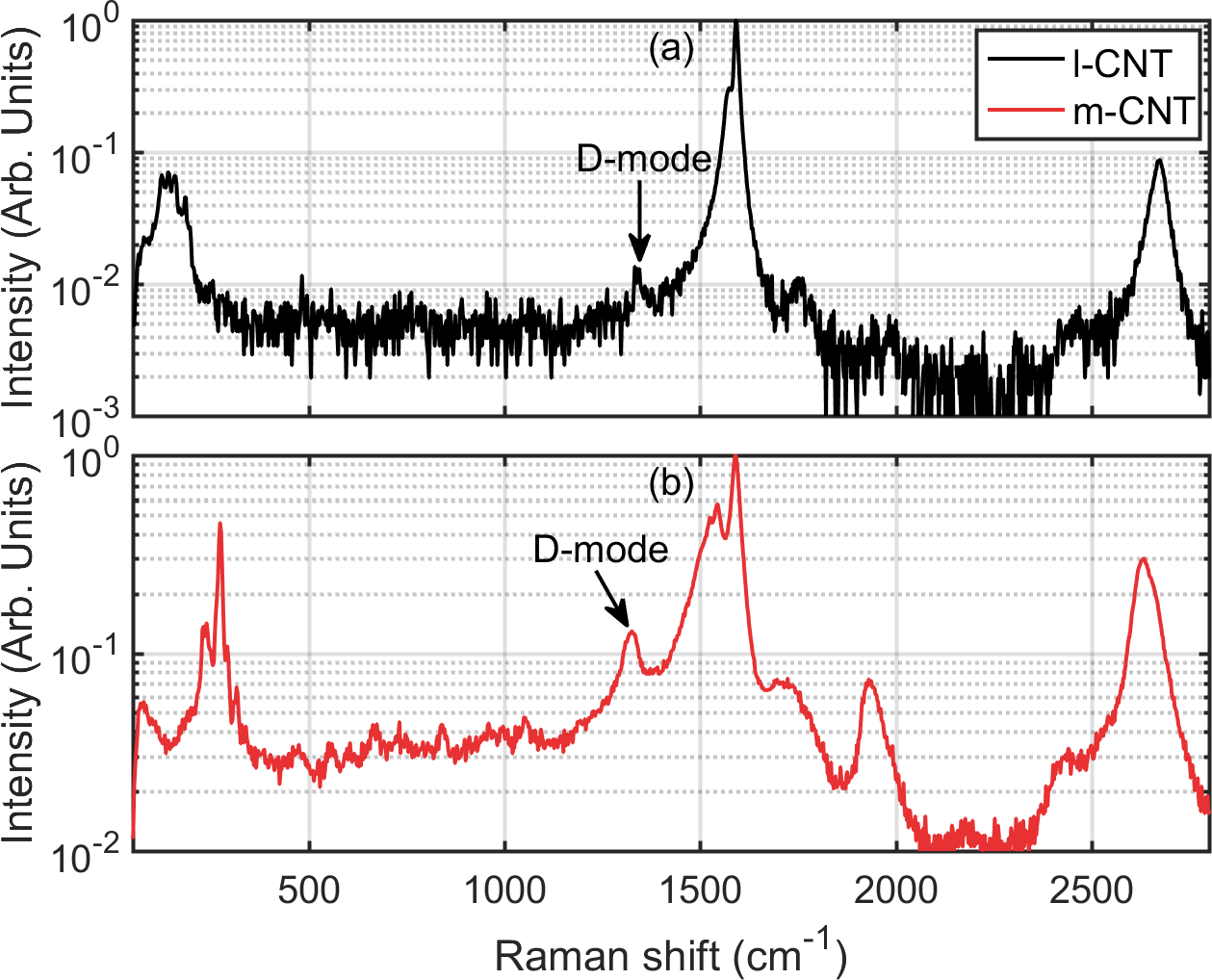}
		\caption{Raman spectra of (a) $l$-CNT and (b) $m$-CNT films. Note the very small D-mode at 1335 cm$^{-1}$, indicating high crystalline quality of the $l$-CNTs.}
		\label{fig:raman}
	\end{figure}

\begin{figure*}[p!]
		\centering
		\begin{subfigure}[t]{0.5\textwidth}
			\centering
			\includegraphics[width=0.99\textwidth]{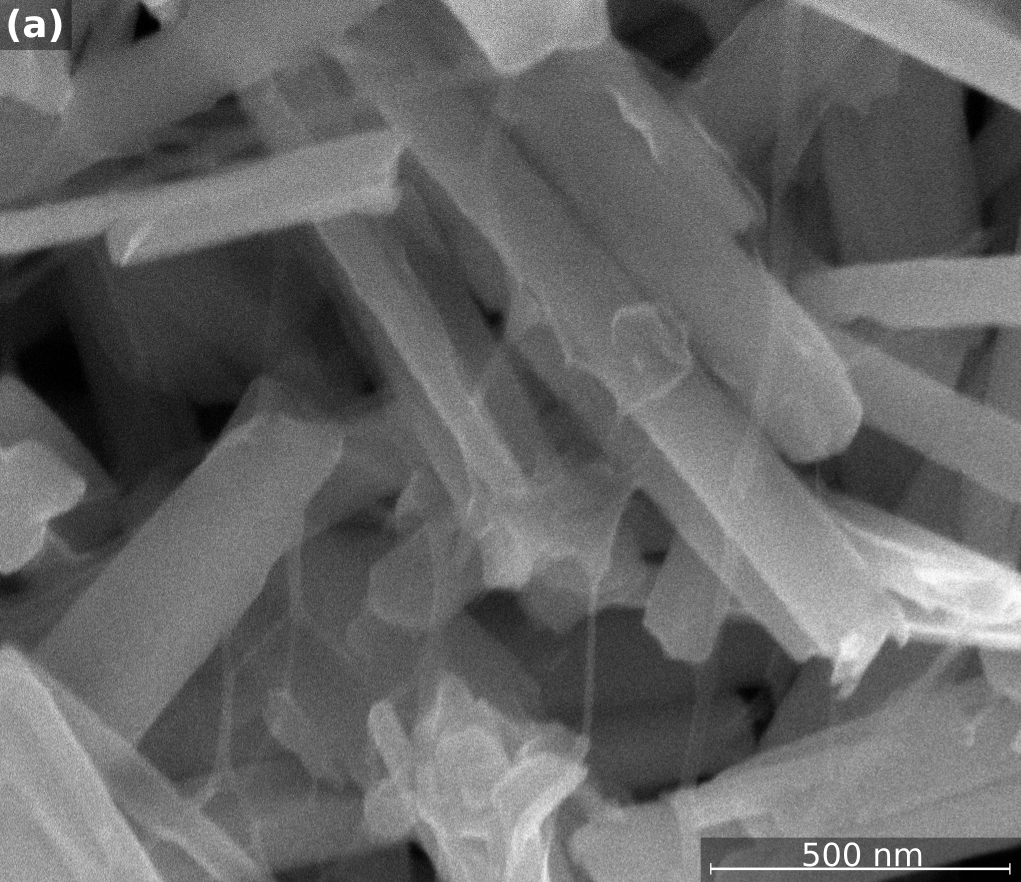}
			\label{fig:m085}
		\end{subfigure}%
		\begin{subfigure}[t]{0.5\textwidth}
			\centering
			\includegraphics[width=0.99\textwidth]{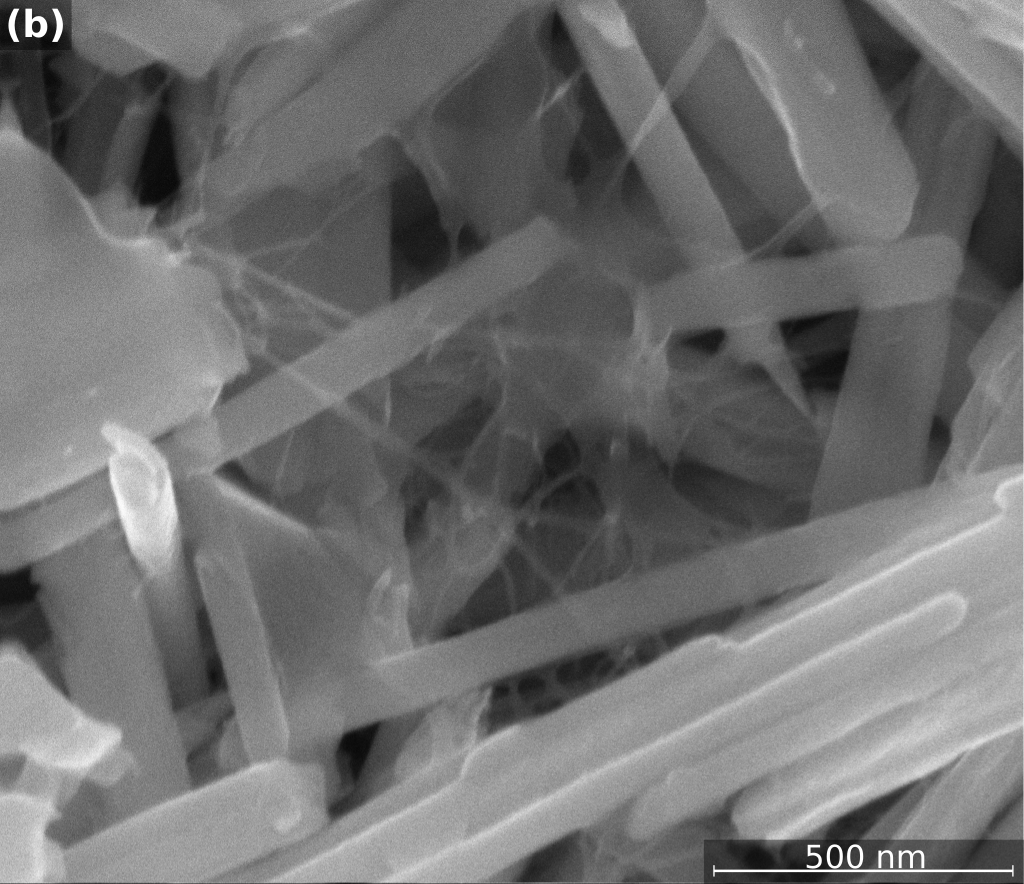}
			\label{fig:m17}
		\end{subfigure}
		\begin{subfigure}[t]{0.5\textwidth}
			\centering
			\includegraphics[width=0.99\textwidth]{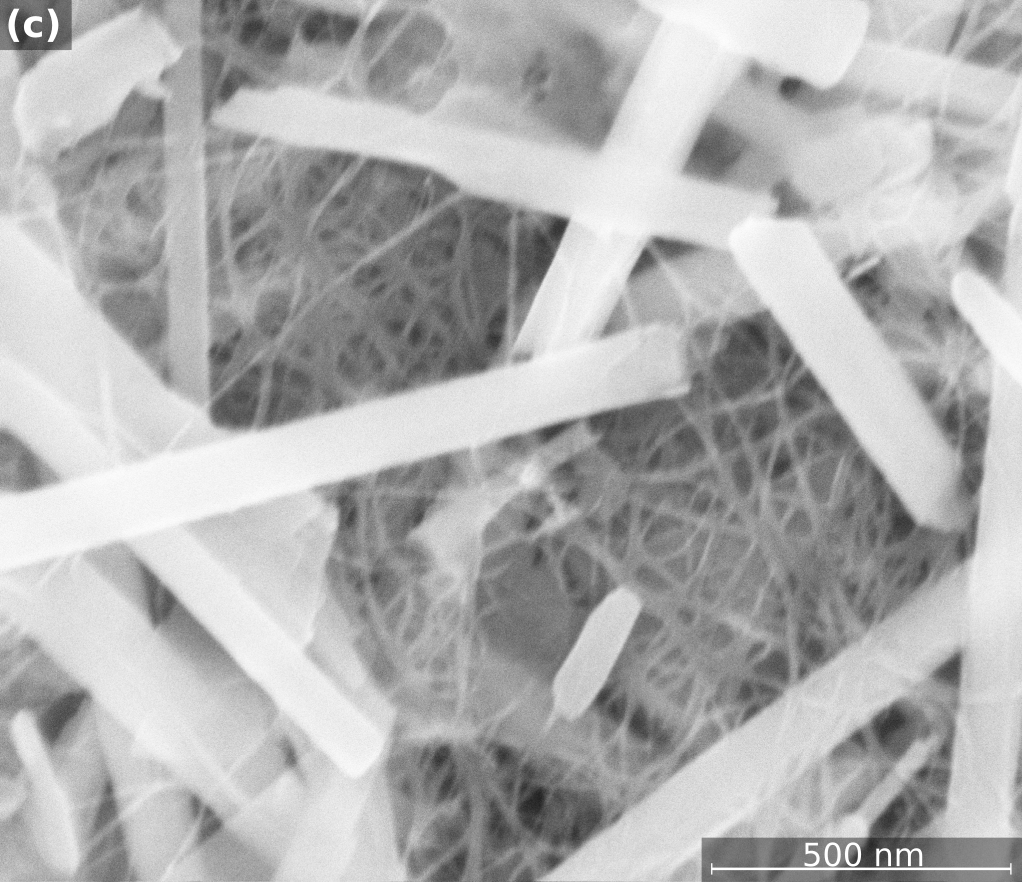}
			\label{fig:m68}
		\end{subfigure}%
		\begin{subfigure}[t]{0.5\textwidth}
			\centering
			\includegraphics[width=0.99\textwidth]{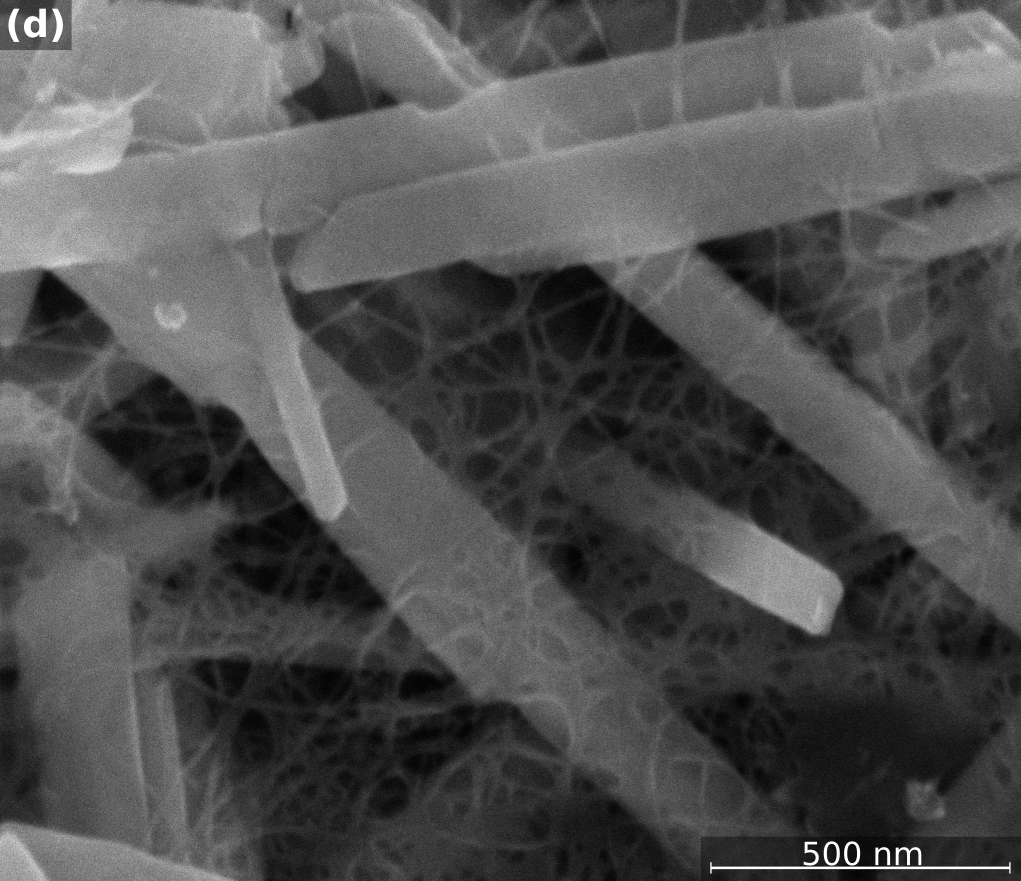}
			\label{fig:X5}
		\end{subfigure}
		\caption{SEM images of samples comprising INTs and (a)-(c) met-CNTs with varying volume fractions of $\phi=0.24\%, 0.49\%, 1.90\%$, respectively, and (d) m-CNTs with volume fraction $\phi=2.4\%$. The thick white tubes are INTs and the thin white curves are the CNT network. Note that the CNTs are well dispersed among the INTs for all CNT volume fractions.}
		\label{fig:SEM-CNT}
	\end{figure*}

	The Raman spectra were obtained using a Raman spectrometer combined with a confocal microscope Nanofinder High End (Tokyo Instruments) at an excitation wavelength of 532nm. Figure \ref*{fig:raman} shows the Raman spectra for pure $m$- and $l$-CNT films. The defect density is estimated from the ratio of the G-band at 1593 cm$^{-1}$, originating from the longitudinal optical phonons of semiconducting CNTs, and the D-band at 1335 cm$^{-1}$ (G/D ratio), which has been found to be a good estimate of the defect density for CNTs with diameters of 1-2 nm, assuming the samples are non-aggregated and without adsorbed molecules \cite{Miyata2011,Jorio2003}. The G/D ratio is approximately 10 for the $m$-CNT films and 100 for the $l$-CNT film, indicating very high crystalline quality in the latter. Therefore, defect scattering can be omitted in the discussion in section 5.2 regarding electron scattering mechanisms in $l$-CNT films.
	
\clearpage
\section{Schematics of experimental setups and procedures}
\begin{figure}[tb]
	\centering
	\includegraphics[width=1\linewidth]{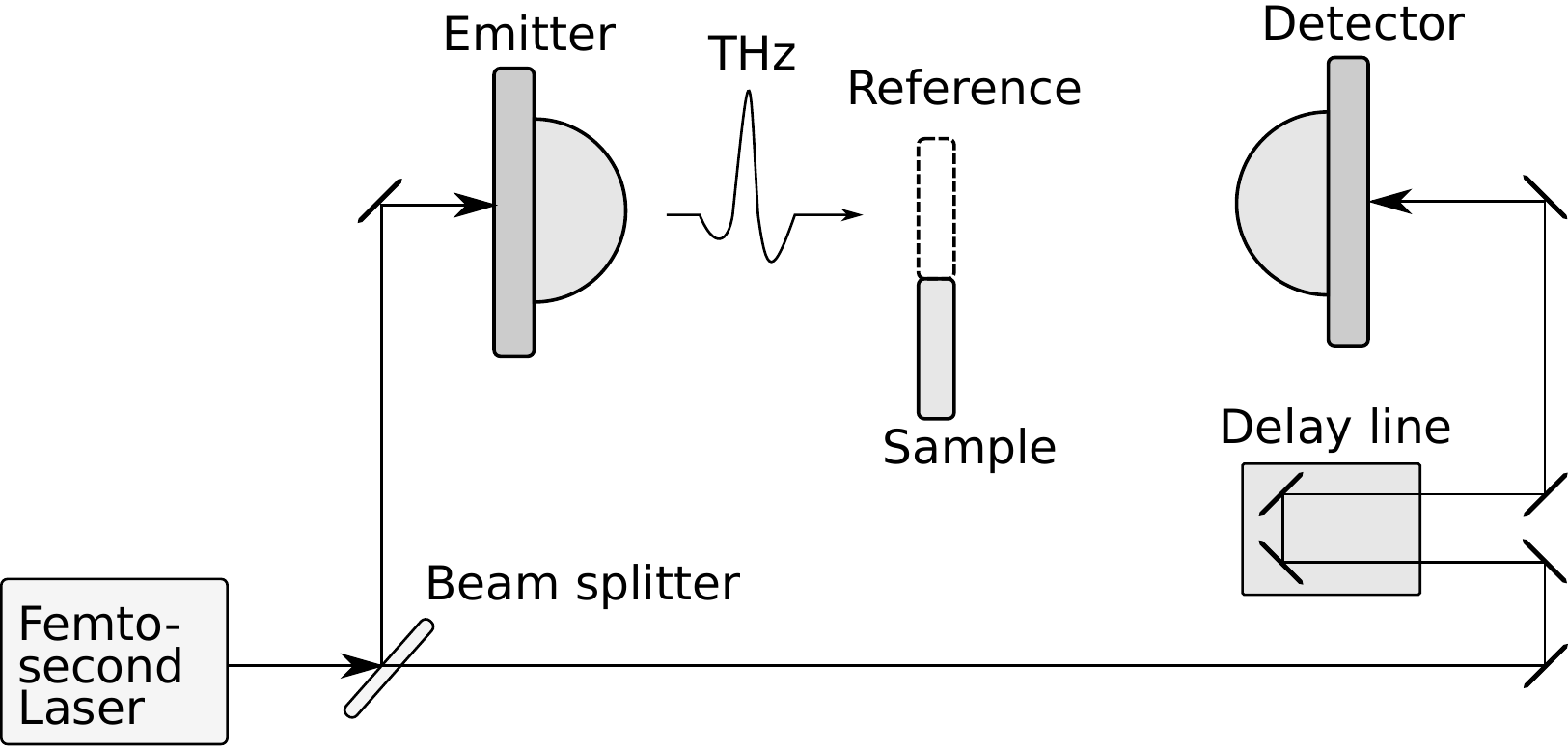}
	\caption{Basic diagram of the terahertz time-domain spectroscopy setup. Femto-second pulses are generated by our laser and split into an emitter and detector path, where they are used to generate and coherently detect THz-pulses. The length of the detector path is controlled by a delay-line. The dielectric properties of a sample is obtained by measuring the THz transmission of the sample and an appropriate reference.}
	\label{fig:thz-tds}
\end{figure}

To investigate the temperature dependence of the THz conductivity in the range 10--300 K, we employed a Terahertz Time-Domain Spectroscopy (THz-TDS) setup with access to a closed cycle helium cryostat (\href{http://www.arscryo.com/}{ARS})\cite{Ulbricht2011}, see figure \ref{fig:thz-tds}.  THz pulses with a bandwidth of 0.3--2 THz were generated and detected by commercially available Photoconductive Antennas (PCAs) from \href{http://www.batop.com/index.html}{Batop} using a 40 MHz, 1064nm, femtosecond fibre laser from \href{http://ekspla.com/}{Ekspla}. By measuring the THz transmission of our samples and an appropriate reference, the conductivity and dielectric function was obtained using equation (3) in the main text.

\begin{figure}[tb]
	\centering
	\includegraphics[width=\linewidth]{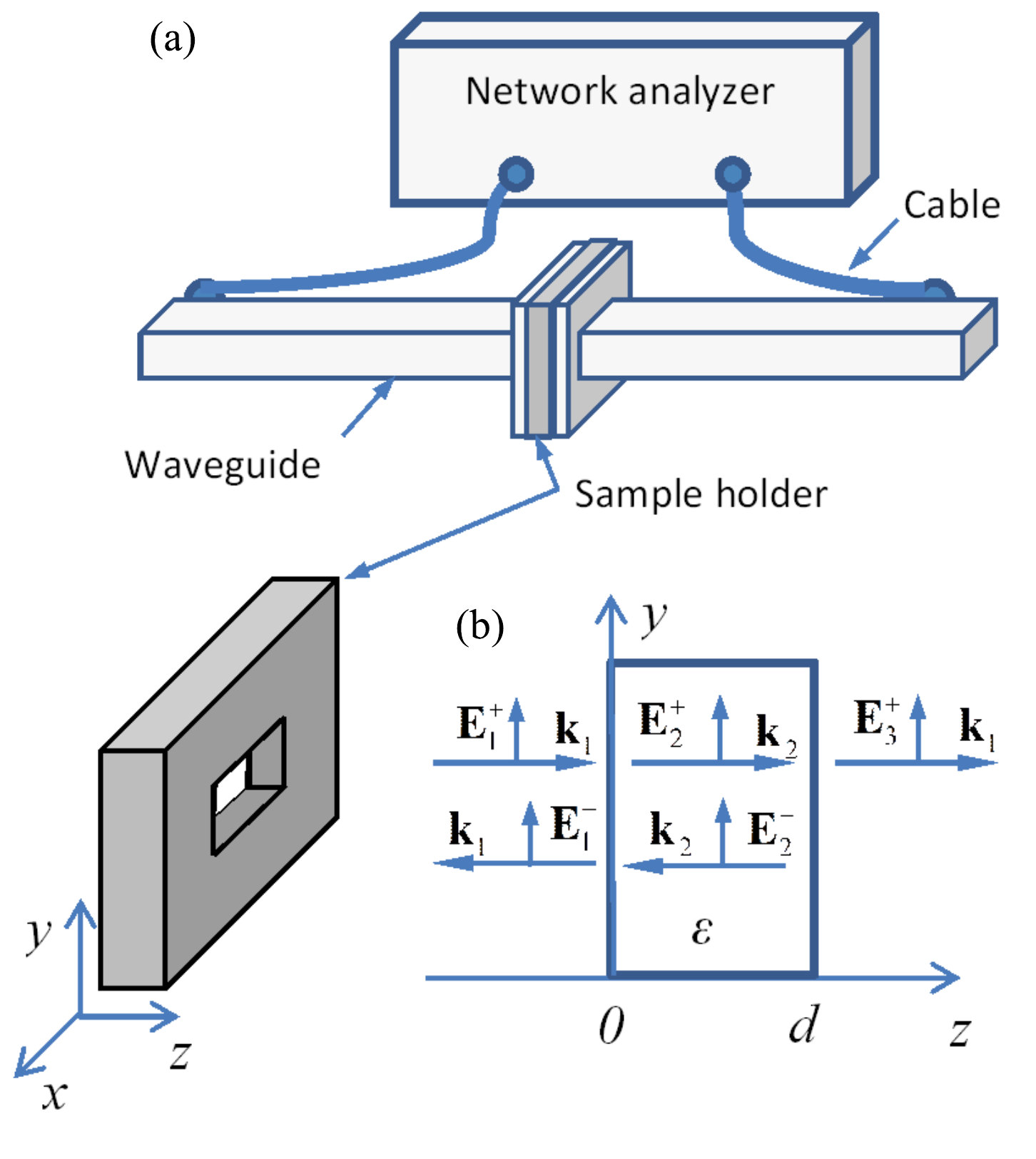}
	\caption{(a) Simplified scheme of the waveguide measurement setup at 30 GHz. (b) The scheme of TE$_{10}$ mode propagation through the thin film of thickness $d$. $\textbf{k}$ and $\textbf{E}$ are wave vector and electric field vector, respectively.}
	\label{GHzsetup}
\end{figure}

The permittivity at 30 GHz was obtained using a scalar network analyzer connected with a rectangular waveguide of dimensions $a \times b$, where $a=7.2$ mm, $b=3.4$ mm (see Fig. \ref{GHzsetup}a).
The CNT films were placed on a 10$\mu$m thick PTFE substrate, which was then placed in the sample holder of the waveguide in a plane orthogonal to the direction of TE$_{10}$ mode propagation. The substrate was assumed to be transparent in the microwave range and therefore it is omitted in further consideration.
Fig. \ref{GHzsetup}b shows the scheme of the TE$_{10}$ mode transmitted through the thin film
of thickness $d$ located at $z\in(0,d)$. The boundary conditions for the propagated field at $z=0$ and $z=d$ are as follows:
\begin{eqnarray}
E_{1}^{+}+E_{1}^{-} = E_{2}^{+}+E_{2}^{-}, \label{eq:GHz1}\\
k_1\left( E_{1}^{+}-E_{1}^{-} \right)=  k_2\left(E_{2}^{+}-E_{2}^{-}\right), \label{eq:GHz2}\\
E_{2}^{+}\exp(ik_2d)+E_{2}^{-}\exp(-ik_2d) = E_{3}^{+}, \label{eq:GHz3}\\
k_2\left[E_{2}^{+}\exp(ik_2d)-E_{2}^{-}\exp(-ik_2d)\right] = k_1 E_{3}^{+},\label{eq:GHz4}
\end{eqnarray}
where 
\begin{eqnarray}
	k_1=\sqrt{(\omega/c)^2-(\pi/a)^2}, \\
	k_2=\sqrt{\varepsilon(\omega/c)^2-(\pi/a)^2}.
\end{eqnarray}
Here $E_{1}^{-}/E_{1}^{+}=S_{11}$ and $E_{3}^{+}/E_{1}^{+}=S_{12}$ are the normalized amplitudes of the reflected and transmitted electric fields measured with the network analyzer. $\varepsilon$ is the complex relative permittivity of the thin film, which is determined from equations (\ref{eq:GHz1}--\ref{eq:GHz4}). More details of the waveguide measurement method can be found in \cite{GHz,Chung2007}. 

	\section*{References}
	\bibliographystyle{iopart-num} % or try abbrvnat or unsrtnat
	\bibliography{literature}